\documentclass[a4paper,12pt]{iopart}
\usepackage{iopams}
\usepackage{setstack}
\usepackage{graphicx}
\usepackage{bm}
\usepackage{epsfig}
\usepackage{hyperref}

\newcommand{\be}{\begin{equation}}
\newcommand{\ee}{\end{equation}}
\newcommand{\bea}{\begin{eqnarray}}
\newcommand{\eea}{\end{eqnarray}}
\newcommand{\nn}{\nonumber}
\newcommand{\la}{\lambda}
\newcommand{\U}{\rm U}

\eqnobysec

\date{\today}

\begin{document}

\title[Lyapunov exponents for products of Ginibre matrices]{Universal distribution of Lyapunov exponents for products of Ginibre matrices}
\author{Gernot Akemann$^{1}$, Zdzislaw Burda$^{2}$ and Mario Kieburg$^{1}$}
\address{$^{(1)}$ Fakult\"at f\"ur Physik, Universit\"at Bielefeld, Postfach 100131, D-33501 Bielefeld, Germany\\
$^{(2)}$ Marian Smoluchowski Institute of Physics, Jagellonian University,
Reymonta 4, 30-059 Krak\'ow, Poland}
\eads{$^{(1)}$ \mailto{akemann@physik.uni-bielefeld.de}, $^{(2)}$ \mailto{zdzislaw.burda@uj.edu.pl}, $^{(3)}$ \mailto{mkieburg@physik.uni-bielefeld.de}}

\begin{abstract}  
Starting from exact analytical results on singular values and complex eigenvalues of products of independent Gaussian complex random $N\times N$ 
matrices also called Ginibre ensemble we rederive the Lyapunov exponents for an infinite product.
We show that for a large number $t$ of product matrices the distribution of each Lyapunov exponent is normal and compute its $t$-dependent variance as well as corrections in a large-$t$ expansion.
Originally Lyapunov exponents are defined for the singular values of the product matrix that represents a linear time evolution. Surprisingly a similar construction for the moduli of the complex eigenvalues yields the very same exponents and normal distributions to leading order. We discuss a general mechanism for $2\times 2$ matrices why the singular values and the radii of complex eigenvalues collapse onto the same value in the large-$t$ limit. Thereby we rederive Newman's triangular law which has a simple interpretation as the radial density of complex eigenvalues in the circular law and study the commutativity of the two limits $t\to\infty$ and $N\to\infty$ on the global and the local scale.
As a mathematical byproduct we show that a particular asymptotic expansion of a Meijer G-function with large index leads to a Gaussian.
\end{abstract}

\section{Introduction}\label{intro}

Lyapunov exponents are useful to study the 
stability of dynamical systems, but they also
play an important role in statistical mechanics of disordered systems, 
localization theory, hidden Markov models and many others areas 
of physics and engineering.

The problem of the determination of Lyapunov exponents is intimately related 
to the asymptotic properties of products of random matrices 
in the limit when the number of factors tends to infinity.
The randomness encoded in these matrices depends 
on the details of the problem in question and it is
usually very difficult to find the exact values
of the exponents. There are however some general theorems that guide the
calculations. For example it is known that the largest Lyapunov exponent of the 
product of a random sequence of matrices generated by a stochastic
process converges almost surely to a limiting deterministic
value in the limit of infinite sequence length. For large but 
finite sequences the largest Lyapunov exponent is a normally 
distributed random variable with the variance inversely 
proportional to the sequence length \cite{fk}.
 
The relevance of products of random matrices to dynamical systems and 
ergodic theory was realized in the sixties \cite{o} and since then the study of 
matrix products has been an active field of research in probability theory
\cite{cknBook}, condensed matter physics, and statistical mechanics 
\cite{blBook,cpvBook,lBook}. 

It was noticed long time ago \cite{d,s} that
products of random matrices naturally arise in the analysis of
disordered systems in statistical mechanics. As an example one can 
think of the transfer matrix formulation of random Ising chains 
\cite{dh,clnp}. In this case the transfer matrices are random matrices. In the thermodynamic limit the free energy density is
given by the largest Lyapunov exponent of the product of transfer matrices. 
Another important physical example is the localization phenomenon 
in electronic systems \cite{a}. In this case the leading Lyapunov  exponent
is related to the inverse localization length \cite{b,hj,t}. 
Other solvable physical models can be found in Yang-Mills theories \cite{lnw}.
In this field unitary transfer matrices in the group $\U(N)$ find applications in calculations of the 
Wilson loop operator for $N\rightarrow \infty$
\cite{jw}.

Products of random matrices have many practical applications in other fields as well.
For instance they arise in calculations of the capacity of a sequence of multiple-input-multiple-output arrays in wireless telecommunication 
\cite{m,akw,aik} and in hidden Markov models  
applied in stochastic inference \cite{cmr}, in time series analysis,  
speech recognition, biological sequence analysis.
In hidden Markov models the Lyapunov exponents correspond to the entropy rates \cite{em,jss}.  Also in image processing~\cite{JaLaJoNi} product matrices play an important role.

The spectrum of Lyapunov exponents gives important information on
the stability and the complexity of dynamical systems \cite{o} and 
their effective information dimension \cite{ky}. 
For this reason a great effort has been made to develop computational methods 
to determine Lyapunov exponents for given theoretical models or to estimate them from 
experimental data. 
Numerical methods are directly based on the analysis of the equation of motion
or measurements of the expansion rates of phase space \cite{bggs1,bggs2}.
Algorithms have been developed for the Lyapunov spectrum from sample time
series \cite{wssv}. Also analytical approximations 
include methods based on the weak disorder expansion \cite{dmp} 
or properties of determinants associated with transfer matrices \cite{p,r}. 

There are only a few models where Lyapunov exponents can be calculated 
exactly. They usually involve products of $2\times2$ matrices
with randomness controlled by a single random parameter 
where the exact expressions result from some model specific 
simplifications which occur during calculations. 
The examples include classical disordered harmonic chains \cite{d,n2}, 
the tight-binding Anderson model \cite{dg,bl}, quantum spin 
chains \cite{nl,fnt,l} and random 
Schr\"odinger operators~\cite{Jens}, see also \cite{blBook,cpvBook,lBook} for reviews.    
Recently a general method has been worked out to derive a scaling form
for the Lyapunov exponents in the continuum limit 
for products of $2\times2$ matrices close to the identity \cite{cltt}
based on the Iwasawa decomposition of SL(2,R) \cite{i}.

An important solvable case where one can calculate the Lyapunov exponents exactly
is the product of identically distributed Gaussian random matrices
with independent, identically distributed (i.i.d.) centered real entries \cite{n}. 
Such matrices are usually called real Ginibre matrices.
This is a special case, first of all because one can analytically derive 
the whole spectrum of Lyapunov exponents $\{\hat{\mu}_1,\ldots,\hat{\mu}_N\}$
for any system size $N$. Second, the calculation 
uncovers a deep connection between the spectrum and the law of large numbers \cite{n}. 
The exponents are exclusively shaped by the statistics of matrix elements and not 
by the matrix structure. In other words the two effects do not mix.  
A second much more recent example where all Lyapunov exponents have been calculated are products of independent Ginibre matrices, where each factor is multiplied by a fixed positive definite matrix \cite{f,k2}. When these constant matrices are equal to the identity the results for the real, complex, and quaternion Ginibre ensembles agree up to a scaling factor $\beta/2$ where $\beta=1,2,4$ is the Dyson index.

The fact that one can derive the whole spectrum is very useful 
for practical purposes since the spectrum can be used to test 
numerical algorithms \cite{bggs1,bggs2,wssv}. Moreover one 
can analytically calculate the limiting law 
for the  distribution of Lyapunov exponents in the limit $N\rightarrow \infty$. 
For the numbers constructed from Lyapunov exponents, that we call in this paper incremental singular values, $\hat{\lambda}_n=\exp[\hat{\mu}_n]$, $n=1,\ldots,N$,
the distribution is given by the triangular law \cite{n}. 

In the present work we further elaborate on the Lyapunov spectrum 
for the product of complex Ginibre matrices. We consider complex Ginibre
matrices that are Gaussian matrices with i.i.d. complex elements. 
We derive an exact form of finite $t$ corrections to
the Lyapunov spectrum, where $t$ is the number of matrices in the product. 
For finite $t$ the Lyapunov exponents are random variables. We calculate
the joint probability distribution for these variables. For
large $t$ it is asymptotically given by a permanent of the product 
of independent Gaussian functions centered at the limiting values. Thereby we determine the widths of the distributions. We also improve this Gaussian approximation by considering another approximation based on the saddle point approximation. The latter approach works even better  for a product of a small number of matrices since it still incorporates asymmetric parts of the individual eigenvalue distributions and to a small extent the original level repulsion.

In addition to the Lyapunov exponents $\hat{\mu}_n$, 
which are related to the singular values of the product matrix, 
one can define the corresponding exponents $\hat{\nu}_n$
for the moduli of the complex eigenvalues. The complex eigenvalue distribution of
the product of Ginibre matrices is rotationally invariant in the complex plane
\cite{bjw,goetze}.
We find that the moduli of the eigenvalues become uncorrelated random
variables in the large-$t$ limit and we determine the form of their joint 
probability distribution.
Surprisingly, the spectrum and the joint probability distribution of these exponents is identical
to that of the Lyapunov exponents, $\hat{\nu}_n=\hat{\mu}_n$ for $n=1,\ldots,N$.

A further consequence of this observation is discussed
in Section \ref{lnl}. The triangular law for Lyapunov exponents corresponding to the singular values found by Isopi and Newman~\cite{ni}
can be understood as the radial distribution of eigenvalues
of the Ginibre matrix. The fundamental reason behind this interpretation is twofold. First, our  insight says that the Lyapunov exponents constructed from the singular values and from the moduli of the eigenvalues agree with each other. Second, Ginibre ensembles belong to
the class of isotropic random matrix ensembles. For those ensembles the sometimes called self-averaging property of the product of isotropic matrices 
\cite{bns,bls} and the Haagerup-Larsen theorem~\cite{hl} are known. These two properties imply that the spectral statistics of a product of independent random matrices is equal to the statistics of the power of a single matrix in the limit of large matrix dimension $N\to\infty$. After taking the root of the product matrix the level density is the one of an ordinary Ginibre matrix which is the circular law for the complex eigenvalues and is equal to the triangular law for the moduli of the eigenvalues.

The paper is organized as follows. In Section \ref{model} we define the
linear evolution given by the product of Ginibre matrices and define the corresponding 
Lyapunov exponents.
In Section \ref{LE_fs} we derive
their joint probability density based on the singular value distribution of the product matrix for finite and large $t$, keeping $N$ finite.
In Section \ref{RE_fs} we compute the joint probability density for 
exponents based on the moduli of complex eigenvalues for finite and large $t$.
In Section \ref{lnl} we discuss the limit $N\rightarrow \infty$
for Lyapunov exponents and show that this limit commutes with the limit $t\to\infty$ on the global scale while it does not commute on the local scale of the mean level spacing.
In Section \ref{ie}
we conjecture the collapse of singular and eigenvalues for general 
isotropic ensembles and exemplify this for $N=2$.
We conclude the paper in Section \ref{conclusion}. In the appendices we recall some identities of Meijer G-functions, compute a particular kind of a Hankel determinant and present some further details of our calculations.

\section{Linear time evolution with Ginibre matrices}\label{model}

Let us consider a linear discrete-time evolution of an $N$-dimensional
system described by $N$ complex degrees of freedom.
The state of the system at time $t$ is given by an $N$-dimensional vector 
$\vec{x}_t$. The state at $t+1$ is related to the state at time $t$ 
by the following linear equation
\begin{equation}
\label{evolution}
\vec{x}_{t+1} = X_{t+1} \vec{x}_t,
\end{equation}
with the evolution operator $X_{t+1}$ represented by an $N\times N$ matrix. 
The total evolution from the initial state 
\begin{equation}
\vec{x}_t = \Pi(t) \vec{x}_0
\end{equation}
is effectively driven by the product matrix
\begin{equation}
\label{Pi}
\Pi(t) \equiv X_t X_{t-1} \cdots X_1.
\end{equation}
Here we are interested in $X_j$'s being i.i.d.
complex non-Hermitian random matrices. In particular we consider the case of Ginibre matrices which centered and Gaussian distributed,
\be
\label{PGin}
d\mu(X_j)=dX_j\exp\left[-\Tr X_j^\dag X_j\right]
\ee
for all $j=1,\ldots,t$. The differential $dX_j$ denotes the product of the differential of all independent matrix elements.
Towards the end of the paper we comment on the evolution for general isotropic random matrices which are defined by the invariance of the probability measure $d\mu(X_j)=d\mu(UX_jV)$ where $U,V\in\U(N)$ are arbitrary unitary matrices. Isotropic matrices are sometimes called bi-unitarily invariant or rotational invariant. Ginibre matrices belong to this class.

We are interested in the large $t$ behavior of the system, approximating a continuous time evolution. 
This behavior is controlled by the Lyapunov exponents which are 
related to the singular values of $\Pi(t)$. Let us denote the real eigenvalues 
of the positive matrix
\begin{equation}
\label{S}
S(t) \equiv \Pi^\dagger(t) \Pi(t)
\end{equation}
by $\{s_n(t)\in\mathbb{R}_+,n=1,\ldots,N\}$. Their square roots $\sqrt{s_n(t)}$ 
correspond to singular values of $\Pi(t)$. Then the Lyapunov exponents are defined as
\begin{equation}
\label{Lydef}
\hat{\mu}_n = \lim_{t\rightarrow \infty} \frac{\ln \hat{s}_n(t)}{2t},
\end{equation}
where $\hat{s}_n(t)$ are the ordered eigenvalues of $S(t)$:
$\hat{s}_1(t) \le \hat{s}_2(t) \le \ldots \le \hat{s}_N(t)$. Throughout this paper
we denote ordered (increasing) sequences
like  $\hat{s}_n$ or $\hat{\mu}_n$ by a hat.
 
In many physical situations the number of time steps in the evolution is large but finite. 
Hence it is interesting to study finite size corrections to the limiting values, 
and the rate of convergence to these values. Thus we want to address the question how this 
limit is realized when $t$ tends to infinity ($t\gg1$). 
Our focus lies on the corresponding quantities for finite $t$
\begin{equation}
\label{ft_L}
\mu_n(t) \equiv \frac{\ln s_n(t)}{2t}\ ,
\end{equation} 
which we call finite $t$ Lyapunov exponents, $\mu_n(t)\in\mathbb{R}$ for $n=1,\ldots,N$. 
In the limit $t\rightarrow \infty$, after ordering, 
they become the standard Lyapunov exponents:  
$\hat{\mu}_n = \lim_{t\rightarrow \infty} \hat{\mu}_n(t)$. 
We look for a probabilistic law that governs the distribution
of the finite $t$ Lyapunov exponents, or equivalently 
their joint probability density $P_N^{(t)}(\mu_1,\ldots,\mu_N)$ for finite 
$t$ and $N$. Given the recent progress on the joint distribution of singular values (and complex eigenvalues) for a finite product of $N\times N$ Ginibre matrices for finite $t$ and $N$ this can be easily calculated, and the limits $t\to\infty$ and subsequently $N\to\infty$ can be taken.


\section{Lyapunov exponents from singular values}\label{LE_fs}

The initial point of our calculations is an exact expression 
for the joint probability distribution of real eigenvalues of the matrix
$S(t)$ (\ref{S}) at finite $N$ and $t$ \cite{akw,aik},
\begin{eqnarray}
\fl P_{N}^{(t)}(s_1,\ldots,s_N)ds_1\cdots ds_N = 
\frac{ds_1\cdots ds_N }{N! \prod_{a=1}^N \Gamma^{t+1}(a)} \Delta_N(s) 
\det\left[G^{t,\,0}_{0,\,t}\left(\left.\mbox{}_{0,\ldots,0,a-1}^{-} \right| \, s_b\right)\right]_{1\leq a,b\leq N},\nonumber\\
\label{Ps}
\end{eqnarray}
where $\Delta_N(s)$ is the Vandermonde determinant
\begin{equation}
\label{vd}
\Delta_N(s) = \det \left[s_a^{b-1}\right]_{1\le a,b \le N} = 
\prod_{1\le a<b\le N} (s_b-s_a).
\end{equation}
The function $G^{t,\,0}_{0,\,t}\left(\left.\mbox{}_{0,\ldots,0,a-1}^{-} \right| \, s\right)$
is a particular case of the Meijer G-function (\ref{Gdef}) whose properties and definition are recalled in \ref{appA}.
As any special function, it possesses many helpful 
properties which facilitate calculations. For simplicity we drop the explicit $t$-dependence of the singular values and of the Lyapunov exponents in the ensuing discussions as it will be clear from the context if $t$ is finite or infinite.

The road map to find the large $t$ asymptotics is the following. In subsection~\ref{sec:cumulant} we find a determinantal representation of the
joint probability distribution of Lyapunov exponents
made of one-point probability distributions. We calculate the moments
of these one-point distributions. The cumulant expansion yields an asymptotic expansion to any order in $1/t$. This result is discussed in detail for large $t$, in subsection~\ref{large-tS}. Moreover, we compare the cumulant expansion with a saddle point approximation which also incorporates a residual level repulsion as well as an asymmetric part of the distributions of the individual Lyapunov exponents.  In subsection~\ref{isv} we come back to the discussion of the corresponding singular values $\exp[\mu_j]$ which we call incremental singular values since they are the average contribution to the total singular value of each single random matrix in the product $\Pi(t)$.

\subsection{Reduction to ``decoupled'' random variables}\label{sec:cumulant}

The joint probability distribution $P_N^{(t)}(\mu_1,\ldots,\mu_N)$ 
for Lyapunov exponents can be directly read off from 
eq. (\ref{Ps}) by the change of variables $s_n\equiv\exp(2t\mu_n)$,
\begin{eqnarray}
\label{Pmu}
\fl P_N^{(t)}(\mu_1,\ldots,\mu_N) d\mu_1\cdots d\mu_N
&=& \frac{(2t)^N d\mu_1\cdots d\mu_N }{N! \prod_{a=1}^N \Gamma^{t+1}(a)}
\det_{1\le a,b \le N} \left[\exp(2tb\mu_a)\right]\\
\fl&&\times
\det_{1\leq a,b\leq N}\left[G^{t,\,0}_{0,\,t}\left(\left.\mbox{}_{0,\ldots,0,a-1}^{-} \right| \, \exp(2t\mu_b)\right)\right] .\nonumber
\end{eqnarray}
The change of variables introduces a Jacobian which yields for each variable $\mu_n$  the exponential factor $ds_n = 2t \e^{t\mu_n} d\mu_n$. These factors 
have been absorbed in the last equation in the Vandermonde determinant 
$\det \left[\exp((b-1)2t\mu_a)\right]$ by replacing $(b-1)\rightarrow b$.
The first determinant in eq.~(\ref{Pmu}) can be expanded as
\begin{equation}
(2t)^N\det_{1\le a,b \le N}  \left[\exp(2tb\mu_a)\right]= 
\sum_{\omega\in S_N} \mathrm{sign}(\omega) \prod_{b=1}^N 2t\exp(2t \omega(b) 
\mu_{b}),
\end{equation}
where $S_N$ denotes the group of permutations of $N$ elements and ``$\mathrm{sign}$'' is the sign function which is $+1$ for even permutations and $-1$ for odd ones.
The factors $2t \exp[2t \omega(b)\mu_b]$ can be absorbed into the second
determinant
\begin{eqnarray}
\fl P_N^{(t)}(\mu_1,\ldots,\mu_N)& =& \frac{1}{{N! \prod_{a=1}^N \Gamma^{t+1}(a)}} \\
\fl&&\times
\sum_{\omega \in S_N} \mathrm{sign}(\omega) 
\det_{1\leq a,b\leq N} \left[2t \exp( 2 t \omega(b)\mu_b) G^{t,\,0}_{0,\,t}\left(\left.\mbox{}_{0,\ldots,0,a-1}^{-} \right|  \exp(2t\mu_b)\right)\right] .\nonumber
\end{eqnarray}
By virtue of eq. (\ref{mGp}) the last expression can be cast into the form
\begin{eqnarray}
\fl P_N^{(t)}(\mu_1,\ldots,\mu_N) &=& \frac{1}{N! \prod_{a=1}^N \Gamma^{t+1}(a)} \\
\fl&&\times
\sum_{\omega \in S_N} \mathrm{sign}(\omega) 
\det_{1\leq a,b\leq N}\left[2t G^{t,\,0}_{0,\,t}\left(\left.\mbox{}_{\omega(b),\ldots,\omega(b),a+\omega(b)-1}^{-} \right| \, \exp(2t\mu_b)\right)\right] .\nonumber
\end{eqnarray}
The skew-symmetry of the determinant under permutations of its rows and columns allows us to absorb the prefactor  $\mathrm{sign}(\omega)$ into the determinant via rearranging the rows. Hence we end up with
\begin{equation}
\label{det}
P_N^{(t)}(\mu_1,\ldots,\mu_N) = 
\frac{1}{N! \prod_{a=1}^N \Gamma^{t+1}(a)} 
\sum_{\omega \in S_N} \det_{1\leq a,b\leq N}\left[ F_{ab}\left(\mu_{\omega(b)}\right)\right]  , 
\end{equation}
where
\begin{equation}
\label{Fab}
F_{ab}(\mu) \equiv 2t G^{t,\,0}_{0,\,t}\left(\left.\mbox{}_{b,\ldots,b,a+b-1}^{-} 
\right| \, \e^{2t\mu}\right) .
\end{equation}
Thus the problem is reduced to the analysis of the 
function $F_{ab}(\mu)$. By construction this function is positive semi-definite. With help of the integral identity~\eref{Ginv}, $F_{ab}$ can be normalized such that the function
\begin{equation}
\label{fF}
f_{ab}(\mu)\equiv \frac{F_{ab}(\mu)}{\int F_{ab}(\mu') d\mu'}=\frac{F_{ab}(\mu)}{\Gamma^{t-1}\left(b\right) \Gamma\left(a+b-1\right)}
\end{equation}
can be interpreted as a probability density for a single random variable.
Replacing $F_{ab}(\mu)$ with its normalized version $f_{ab}(\mu)$ the
joint probability distribution reads
\begin{equation}
\label{Pf}
\fl P_N^{(t)}(\mu_1,\ldots,\mu_N) = \frac{1}{N! \prod_{a=1}^N \Gamma^2(a)} 
\sum_{\omega \in S_N}
\det_{1\leq a,b\leq N}\left[\Gamma(a+b-1) f_{ab}(\mu_{\omega(b)})
\right]  . 
\end{equation}
In passing from eq. (\ref{det}) to eq. (\ref{Pf}) we have pulled the
factor $\prod_{a=1}^N \Gamma^{t-1}(a)$ out of the determinant.
This factor cancels the corresponding prefactor in eq.~(\ref{det}) 
leaving the product of the second powers in front of the determinant in eq.~(\ref{Pf}).

Using the cumulant expansion
we argue in the next subsection that the probability densities $f_{ab}(\mu)$ can
be approximated by Gaussian functions in the limit $t\rightarrow \infty$. 
Therefore let us define the moment generating function  
\begin{equation}
\label{mdef}
M_{ab}(\vartheta)\equiv\int_{-\infty}^{+\infty} d\mu \; \exp(\mu \vartheta) f_{ab}(\mu) =
\sum_{n=0}^\infty \frac{\vartheta^n}{n!} \langle \mu^n \rangle_{ab} \ .
\end{equation}
where $\langle \mu^n \rangle_{ab}\equiv\int_{-\infty}^{+\infty} d\mu f_{ab}(\mu) \mu^n$ are the moments.
This moment generating function can be calculated with help of eq.~\eref{Ginv},
\begin{equation}
\label{Mab}
M_{ab}(\vartheta) = 
\frac{\Gamma^{t-1}\left(b+\vartheta/(2t)\right)
\Gamma\left(a+b-1+\vartheta/(2t)\right)}
{\Gamma^{t-1}\left(b\right) 
\Gamma\left(a+b-1\right)} .
\end{equation}
The expansion in $\vartheta$ at $\vartheta=0$ yields the moments
$\langle \mu^n \rangle_{ab}$.
The logarithm of the moment generating function is the cumulant generating 
function
\begin{equation}
\label{g}
\fl g_{ab}(\vartheta) \equiv \ln \left( M_{ab}(\vartheta)\right) =
(t-1) \ln \left( \frac{\Gamma\left(b+\vartheta/(2t)\right)}{\Gamma\left(b\right)}\right)
+
\ln \left( \frac{\Gamma\left(a+b-1+\vartheta/(2t)\right)}
{\Gamma\left(a+b-1\right)} \right) .
\end{equation}
The coefficients of the corresponding Taylor series of $g_{ab}(\vartheta)$ at $\vartheta=0$  are the cumulants 
$\kappa^{(n)}_{ab}$,
\begin{equation}
\label{cumulant_g}
\fl g_{ab}(\vartheta) \equiv \sum_{n=1}^\infty \frac{\vartheta^n }{n!} \kappa^{(n)}_{ab} =
\sum_{n=1}^\infty \frac{\vartheta^n}{(2t)^{n-1}n!} \left(\frac{\psi^{(n-1)}(b)}{2}
+ \frac{\psi^{(n-1)}(a+b-1)-\psi^{(n-1)}(b)}{2t} \right).
\end{equation}
Hereby we employed the definition of the digamma function and its derivatives,
\begin{equation}
\fl\psi(x) = \frac{d}{dx} \ln \Gamma(x),\quad \psi^{(n)}(x)= \frac{d^n}{dx^n}\psi(x)\quad \left(\psi^{(0)}(x)\equiv \psi(x),\ \psi^{(1)}(x)\equiv \psi'(x)\right).
\end{equation}
The first cumulant (=first moment) corresponds to the mean value 
$\kappa^{(1)}_{ab}=\langle \mu \rangle_{ab}
=\int d\mu f_{ab}(\mu) \mu$ and is equal to
\begin{equation}
\label{mab}
m_{ab}\equiv\kappa^{(1)}_{ab}= \frac{\psi(b)}{2} + 
\frac{\psi(a+b-1)-\psi(b)}{2t} \ .
\end{equation}
The second cumulant corresponds to the variance 
$\kappa^{(2)}_{ab}=\int d\mu f_{ab}(\mu) \left(\mu-m_{ab}\right)^2$ and takes the value
\begin{equation}
\label{sab}
(\sigma_{ab})^2\equiv \kappa^{(2)}_{ab} = \frac{1}{2t}
\left(\frac{\psi'(b)}{2} + \frac{\psi'(a+b-1)-\psi'(b)}{2t}\right) .
\end{equation}
We emphasize that so far all results are exact for finite $t$.

\subsection{Large $t$ limit}\label{large-tS}

We apply the standard argument based on the analysis of the large-$t$ behavior
of cumulants to show that $f_{ab}(\mu)$ can be approximated 
by a Gaussian function for large $t$. Thereby we have first to center the distribution $f_{ab}(\mu)$ and normalize its second moment. The exact limit $t\to\infty$ will yield a Gaussian. This limit justifies to replace $f_{ab}(\mu)$ by a Gaussian centered at $m_{ab}$ and with the standard deviation $\sigma_{ab}$.

For this purpose we define the standardized random
variable $\mu_*\equiv(\mu-m_{ab})/\sigma_{ab}$. Thereby we denote standardized quantities by $*$ in this and the next section. The random variable $\mu_*$ is distributed as $f_{*ab}(\mu_*) \equiv \sigma_{ab} f(\mu_* \sigma_{ab} + m_{ab})$. 
The same notation is applied for cumulants.
By construction, the standardized mean is
$m_{*ab}=0$ and the standardized variance is $\sigma_{*ab}=1$.
The higher standardized cumulants are
\begin{equation}
\label{kappan}
\kappa^{(n)}_{*ab} \equiv \frac{\kappa^{(n)}_{ab}}{(\sigma_{ab})^n} \sim
t^{1-n/2} \longrightarrow 0 \ , \ n=3,4,\ldots
\end{equation}
They tend to zero when $t$ goes to infinity. Therefore the standardized cumulant generating function is in the limit $t\to\infty$,
\begin{equation}
\lim_{t\to\infty}g_{*ab}(\vartheta) = \frac{1}{2}\vartheta^2.
\end{equation}
By analytic continuation to imaginary values
$\vartheta = \imath \omega$ we get
$\lim_{t\to\infty}g_{*ab}(\imath\omega) = -\omega^2/2$
and hence $\lim_{t\to\infty}M_{*ab}(\imath\omega) = \exp(-\omega^2/2)$. The inverse Fourier
transform for the moment generating function yields the limit
\begin{equation}
\lim_{t\to\infty}f_{*ab}(\mu) = \frac{1}{\sqrt{2\pi}}\ \exp\left[-\frac{\mu^2}{2}\right] .
\end{equation}
Inverting the process of standardization 
$f_{ab}(\mu) = \sigma^{-1}_{ab} f_{*ab}\left((\mu-m_{ab})/\sigma_{ab}\right)$
we get the following asymptotic expansion
\begin{equation}
\label{gauss}
\fl f_{ab}(\mu) =
 \frac{2t\ G^{t,\,0}_{0,\,t}\left(\left.\mbox{}_{b,\ldots,b,a+b-1}^{-} 
\right| \, \exp(2t\mu)\right) 
}{\Gamma^{t-1}\left(b\right) \Gamma\left(a+b-1\right)}
\overset{t\gg1}{\approx} \frac{1}{\sqrt{2\pi(\sigma_{ab})^2}} 
\exp\left( -\frac{(\mu-m_{ab})^2}{2(\sigma_{ab})^2}\right) \ ,
\end{equation}
with $m_{ab}$ and $\sigma_{ab}$ given by eqs. (\ref{mab}) and (\ref{sab}).
In other words, for large $t$ we can replace $f_{ab}(\mu)$ in (\ref{Pf})
by the Gaussian function eq. (\ref{gauss}). Here we have also reinserted the definition of $f_{ab}(\mu)$ from eqs. (\ref{fF}) and (\ref{Fab}) in order to stress that this is the first main result of this section, namely the asymptotic expansion of a Meijer G-function in the double scaling limit of large argument and large index. We are not aware of such a result in the literature. In particular it is different from the well-known large argument expansion, cf. \cite{fields}.

The expression~(\ref{Pf}) can be further simplified for large $t\gg1$ since
the mean value $m_{ab} \longrightarrow m_{b}$, cf. eq. (\ref{mab}), 
and the variance $(\sigma_{ab})^2\longrightarrow (\sigma_b)^2$, cf. eq. (\ref{sab}), asymptotically depend on a single index 
\begin{equation}
\label{single_index}
 m_{b} \equiv \frac{\psi(b)}{2} \ , \ \sigma_b^2 \equiv  \frac{\psi'(b)}{4t} 
\end{equation}
and hence $f_{ab}(\mu) \longrightarrow f_b(\mu)$ with 
\begin{equation}
\label{single_index_peak}
f_b(\mu) \equiv  \frac{1}{\sqrt{2\pi\sigma^2_{b}}} 
\exp\left( -\frac{(\mu-m_{b})^2}{2\sigma_{b}^2}\right) ,
\end{equation}
which was known for $b=N$ \cite{cknBook}.
Since these functions are independent of the index $a$,
after replacing $f_{ab}(\mu_{\varepsilon(b)})$ by $f_{b}(\mu_{\varepsilon(b)})$ we can pull the factors $f_{b}(\mu_{\varepsilon(b)})$ out the determinant in eq. (\ref{Pf}). This yields  
\begin{eqnarray}
P_N^{(t)}(\mu_1,\ldots,\mu_N)& \overset{t\gg1}{\approx} &\frac{\det_{1\leq a,b\leq N}\left[\Gamma(a+b-1)\right]}{{N! \prod_{a=1}^N \Gamma^2(a)}} 
\sum_{\varepsilon \in S_N}\prod_{b=1}^N f_b\left(\mu_{\varepsilon(b)}\right)  \nonumber\\
&=&\frac{1}{N!}\mathrm{per}_{1\le a,b\le N}\left[f_b(\mu_a)\right] .
\end{eqnarray}
Here the sum over permutations without signs is equal to the definition of the permanent,  
$\mathrm{per}_{1\le a,b\le N}\left[f_b(\mu_a)\right]$.
The prefactor simplifies to $1/N!$ since 
\begin{equation}
\label{Hankel}
\det_{1\leq a,b\leq N}\left[\Gamma(a+b-1)\right] = \prod_{a=1}^N\Gamma^2(a) \ ,
\end{equation}
as recalled in \ref{appB}.

Let us state the main result of this section in its explicit form which is the joint probability distribution for large $t$,
\begin{eqnarray}
P_N^{(t)}(\mu_1,\ldots,\mu_N)&\overset{t\gg1}{\approx}&
\frac{1}{N!} \mathrm{per}_{1\le a,b\le N}\left[\sqrt{\frac{2t}{\pi\psi'(b)}} 
\exp\left( -t\frac{(2\mu_a-\psi(b))^2}{2\psi'(b)}\right)\right] \nonumber\\
& \equiv&
P_N(\mu_1,\ldots,\mu_N).
\label{leading}
\end{eqnarray}
The limiting joint probability distribution
sustains its invariance under permutations of the indices, $P_{N}(\mu_1,\ldots,\mu_N)=P_{N}(\mu_{\omega(1)},\ldots,\mu_{\omega(N)})$. 
More explicitly, the joint probability density is a symmetrized product of one-point functions or densities, which means in physical language that it describes a system of $N$ independent, non-interacting, indistinguishable bosons.
Starting from the determinantal process of the singular values the appearance of a permanent is somewhat surprising, whereas it quite naturally arises for complex eigenvalues after integrating over the angles, see e.g. in \cite{as,ais}. We will come back to this point at the end of section \ref{RE_fs}.

Note that the dependence of $P_N^{(t)}$ on $t$ appears only through the widths of the Gaussian peaks.
Their positions are independent of $t$ in this approximation. 

The density defined as 
\be
\rho_{N}(\mu) \equiv \int d\mu_2 \ldots d\mu_N P_N(\mu,\mu_2,\ldots,\mu_N)
\label{rhodef}
\ee
is in our case
\begin{equation}
\label{one-point}
\rho_{N}(\mu) = \frac{1}{N} \sum_{b=1}^N f_b(\mu) = 
\frac{1}{N} \sum_{b=1}^N \frac{1}{\sqrt{2\pi\sigma^2_b}} 
\exp\left( -\frac{(\mu-m_{b})^2}{2\sigma_{b}^2}\right) . 
\end{equation}  
When $t$ increases the peaks become more narrow and, eventually in the limit $t\rightarrow \infty$,
the Gaussian peaks turn into Dirac delta functions and we recover
the deterministic laws \cite{n,f}
for the Lyapunov exponents $\hat{\mu}_b=\psi(b)/2$,
\begin{equation}
\label{deterministic}
\lim_{t\to\infty}\rho_{N}(\mu) = \frac{1}{N} \sum_{b=1}^N 
\delta\left(\mu - \frac{\psi(b)}{2}\right) . 
\end{equation}  
Employing Newman's argument~\cite{n} one can show that the positions of the peaks for general Dyson index $\beta=1,2,4$ are given by $\psi(\beta b/2)/2$ with $b\in\mathbb{N}$. Thus the positions we calculated fit into the results obtained for products of real Ginibre matrices by Newman~\cite{n} and agree with the more general recent result by Forrester~\cite{f} who considered complex Ginibre matrices multiplied by a fixed positive definitive matrix. Forrester's work was extended by Kargin \cite{k2} to $\beta=1,4$.
Let us emphasize that our  result~(\ref{one-point}) gives
finite-$t$ corrections to this deterministic law. 
Moreover we stress that the same limit has a corresponding consequence for the Meijer G-functions for the individual peaks, when taking the limit $t\to\infty$,
\be
\lim_{t\to\infty}
 \frac{
2t G^{t,\,0}_{0,\,t}\left(\left.\mbox{}_{b,\ldots,b,a+b-1}^{-} 
\right| \, \exp(2t\mu)\right) 
}{\Gamma^{t-1}\left(b\right) \Gamma\left(a+b-1\right)}
= \delta\left(\mu - \frac{\psi(b)}{2}\right) 
\ee
and
\be
\fl \lim_{t\to\infty}
 \frac{
2t\sigma_{ab} G^{t,\,0}_{0,\,t}\left(\left.\mbox{}_{b,\ldots,b,a+b-1}^{-} 
\right| \, \exp(2t(\sigma_{ab}\mu_*+m_{ab})\right) 
}{\Gamma^{t-1}\left(b\right) \Gamma\left(a+b-1\right)}
= \frac{1}{\sqrt{2\pi}}\ \exp\left[-\frac{\mu^2}{2}\right] .
\ee

Already for finite but sufficiently large $t$ when the 
peaks cease to overlap, each Gaussian peak $f_b(\mu)$, see 
eq. (\ref{single_index_peak}), can be identified as a finite size distribution
of the $(N-b+1)$-th largest Lyapunov exponent $\hat{\mu}_b$. 
Due to the recursion $\psi(b+1)=\psi(b)+1/b$ the distance between neighboring peaks is 
$m_{b+1}-m_{b} = 1/(2b)$ and the sum of their widths is 
$\sigma_{b+1} + \sigma_{b} \approx 1/\sqrt{bt}$. So the peaks
separate when $(m_{b+1}-m_{b})\gg (\sigma_{b+1} + \sigma_{b})$ implying $t \gg 4 b$. Thus, for the system with $N$ 
degrees of freedom all peaks get separated for $t \gg 4 N$. 
Note that the positions $m_b$ and the widths $\sigma_b$
are independent of $N$. When $N$ increases, just new peaks appear in the distribution
while the old ones neither change in shape nor shift their positions.

\begin{figure}
\centerline{\includegraphics[width=1\textwidth]{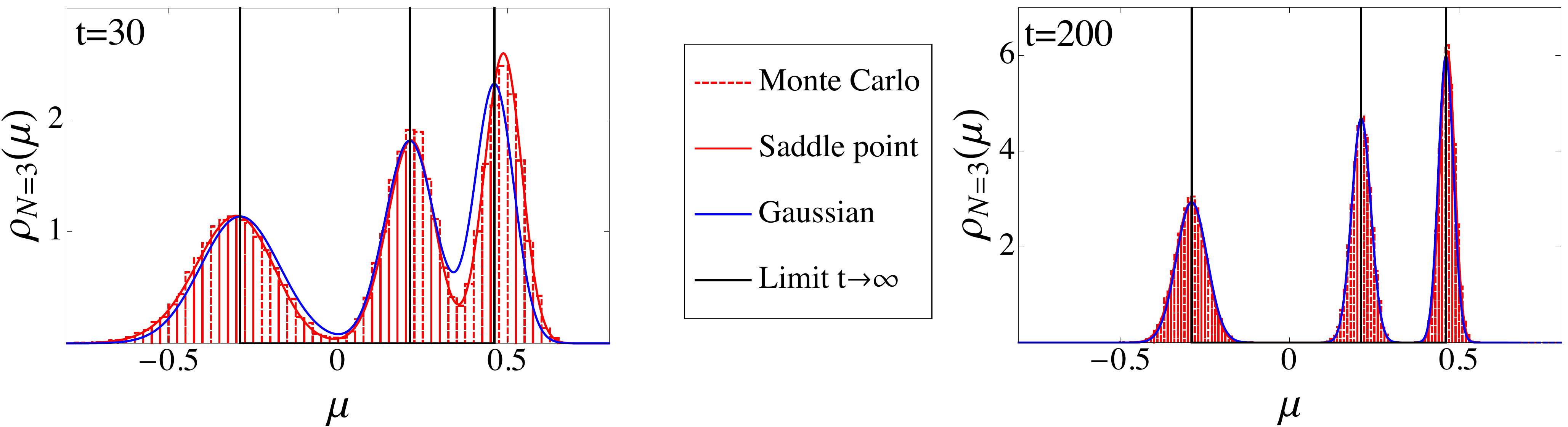}}
\caption{Comparison of the density of Lyapunov exponents $\rho_{N=3}(\mu)$  
given in the Gaussian  approximation~(\ref{one-point}) (blue curve), in the saddle point approximation~(\ref{Gaussapprox}) (red curve) and
generated by Monte Carlo simulations (red histogram, ensemble size = $10000$ product matrices).  We consider products of $t=30$ (left plot) and of $t=200$ (right plot) matrices. The peaks (black vertical lines) are located at $\mu=\psi(b)/2$ for $b=1,2,3$, which are approximately equal to $\{-0.29,0.21,0.46\}$. Note that the Gaussian approximation yields a good agreement only
if $t$ is large enough. But even then the deviations become visible
for larger Lyapunov exponents. The saddle point approximation works better
since it incorporates lower order corrections. Nevertheless also the saddle point approximation has its limits explaining the small, but remaining deviations from the numerics.
\label{leading_vs_1t}}
\end{figure}

Let us study the quality of the approximation that has led us to eq.~(\ref{leading}).
In the derivation of the asymptotic form (\ref{leading}) 
for large $t$ we used the fact that the functions $f_{ab}(\mu)$ 
can be approximated by Gaussian functions 
(\ref{gauss}) and that their mean values $m_{ab}$ and their variances 
$(\sigma_{ab})^2$ asymptotically only depend on
a single index $b$, see eq.~(\ref{single_index}), if one neglects
$1/t$ terms. The 1/t terms have a twofold effect on the shape of the density.
First, the positions and widths of the peaks solely resulting from the single random variable distributions $f_b(\mu)$ weakly 
dependent on $t$. Second a repulsion between peaks is introduced due to the  determinant in eq. (\ref{det}). 
We illustrate these two effects in Fig.~\ref{leading_vs_1t} for the level density where we compare 
the asymptotic formula~(\ref{one-point}) and a saddle point approximation of $f_{ab}(\mu)$ for the inverse Fourier transform of the moment generating function~\eref{mdef},
\begin{eqnarray}
 \fl f_{ab}(\mu)&\overset{t\gg1}{\approx}&\sqrt{\frac{2t}{\pi\psi'(\vartheta_0(\mu))}}\frac{\Gamma^{t-1}(\vartheta_0(\mu))\Gamma(a-1+\vartheta_0(\mu))}{\Gamma^{t-1}(b)\Gamma(a+b-1)}\exp[-2t\mu(\vartheta_0(\mu)-b)] \nonumber\\
 \fl&\equiv&\frac{h_{ab}(\mu)}{\Gamma(a+b-1)},\label{saddlepointapp}
\end{eqnarray}
with
\begin{equation}\label{deftheta0}
 \vartheta_0(\mu)=\int_0^\infty dy\Theta(2\mu-\psi(y))
\end{equation}
and $\Theta$ being the Heaviside function. This approximation is derived in \ref{appC}. Note that in the large $t$ limit the distribution $h_{ab}$ indeed becomes the Gaussian~\eref{single_index_peak} and independent of the index $a$. The level density in the approximation~\eref{saddlepointapp} is
\begin{eqnarray}
\fl \rho_N^{(t,{\rm Saddle})}(\mu) &\equiv&\frac{1}{N\prod_{a=1}^N\Gamma(a)}\sum_{j,l=1}^N(-1)^{l+j}\underset{a\neq j,b\neq l}{\underset{1\leq a,b\leq N}{{\det}}}\left[\Gamma(a+b-1)\right]
h_{jl}(\mu)\nonumber\\
\fl&=&\frac{1}{N}\sum_{j,l=1}^N(-1)^{l+j}\left(\sum_{k=0}^{N-1}\frac{(k!)^2}{\Gamma(k-j+2)\Gamma(k-l+2)}\right)\frac{h_{jl}(\mu)}{[(j-1)!(l-1)!]^2} .
\label{Gaussapprox}
\end{eqnarray}
Hereby we integrated over all but one Lyapunov exponents, $\mu_{1},\ldots,\mu_{N-1}$, and we expanded the determinant~\eref{Pf} in the columns and rows where the remaining distribution $f_{ab}(\mu)\approx h_{ab}(\mu)$ stands. Note that $f_{ab}$ as well as $h_{ab}$ are normalized. The cofactor of the Hankel determinant~\eref{Hankel} is calculated in \ref{appB}.

The main conclusion from
the comparison in Fig.~\ref{leading_vs_1t} is that the corrections do not have any significant effect 
on the shape of the distribution when the peaks are separated. In particular for the smallest singular values this requirement is often satisfied. Nevertheless the corrections can become quite important for $t\approx N$ up to $10N$ in which case the saddle point approximation~\eref{saddlepointapp} is better suited. For the largest eigenvalues the effect of these corrections is the strongest.

In Fig.~\ref{leading_vs_1t} we compare our analytical results with Monte-Carlo
simulations for $3\times3$ product matrices, too. Within the numerical accuracy the agreement is quite good for the Gaussian approximation~\eref{one-point} for $t=200$ and becomes better for the saddle point approximation~\eref{Gaussapprox} already at $t=30$.

\subsection{Incremental singular values} \label{isv}

\begin{figure}
\centerline{\includegraphics[width=0.9\textwidth]{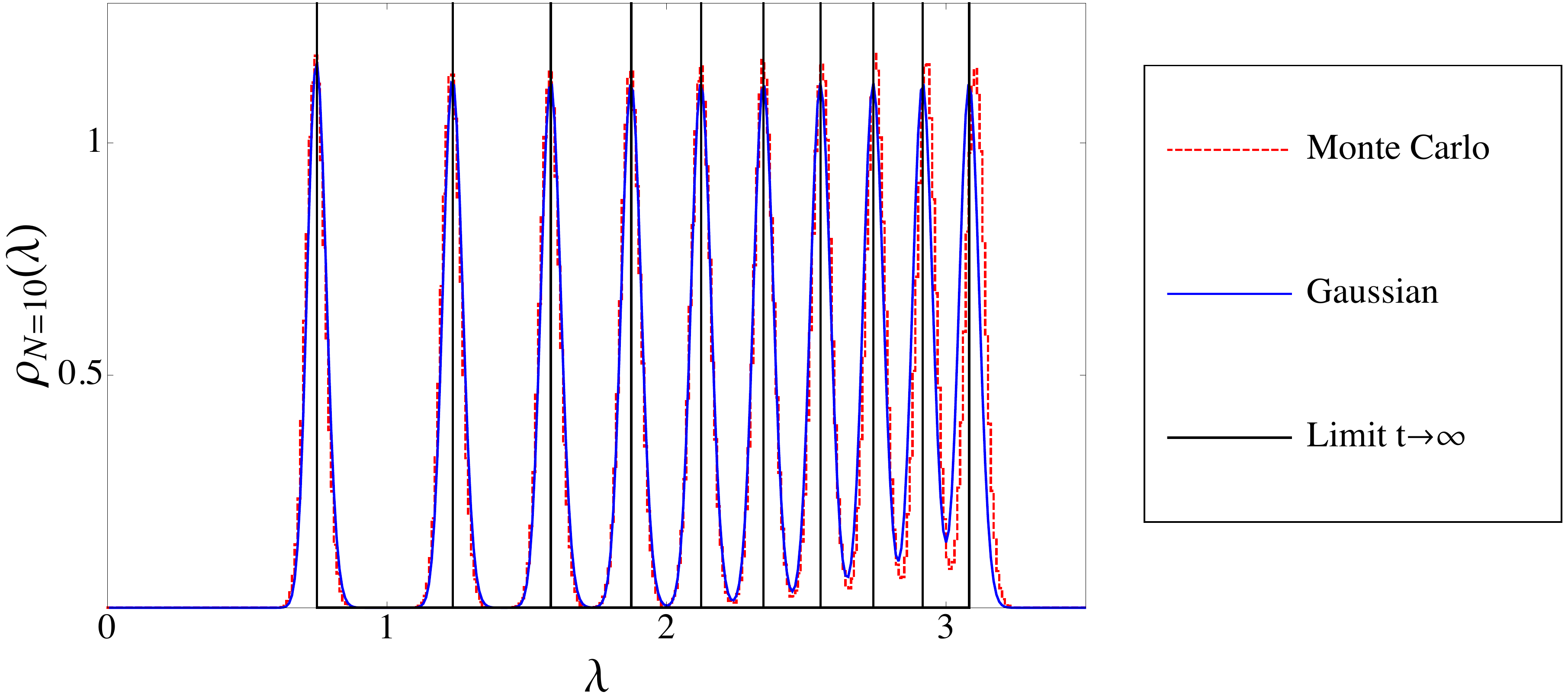}}
\caption{Shown is the comparison of the analytical prediction~(\ref{rhoincr}) (blue curve) and Monte-Carlo simulations (red dashed histogram, ensemble size = $1000$ product matrices) of the density of incremental singular values $\rho_{N=10}(\lambda)$.
The number of matrices multiplied is $t=200$. The sharp peaks appearing for $t\to\infty$ are shown by black vertical lines at the positions $\exp[\psi(b)/2]$, $b=1,\ldots,10$. The deviation increases for larger singular values as expected since the overlap of the peaks becomes stronger.
\label{esv_histograms}}
\end{figure}

We close this section by going back to the singular values because 
in some physical situations it is more convenient to use them
rather than Lyapunov exponents. 
Consider the $t$-th root of the matrix $S(t)$,
\begin{equation}
\label{lambdaPi}
\Lambda(t) = \left( \Pi^\dagger(t) \Pi(t)\right)^{1/(2t)},
\end{equation}
in contrast to eq.~(\ref{S}).  We define  incremental singular values as
\begin{equation}
\label{ft_sv}
\lambda_n(t) \equiv \exp(\mu_n(t)) = s_n^{1/(2t)}(t)\ ,
\end{equation}
which correspond to the real positive eigenvalues of the matrix $\Lambda(t)$.
Intuitively, the  incremental singular values  $\lambda_n(t)$ give the typical incremental contraction or expansion 
factors for the configuration space under a single average time step of the evolution. 
Of course they contain exactly the same information as the Lyapunov exponents.
Their joint probability distribution is obtained from 
that for the Lyapunov exponents by the simple change of variables in eq. (\ref{ft_sv})
inserted in eq. (\ref{Pmu}). Using eq. (\ref{Pf}) this gives
\bea
\fl P_N^{(t)}(\la_1,\ldots,\la_N)d\la_1 \cdots d\la_N &=&
\lambda_1^{-1} \cdots \lambda_N^{-1} P_N^{(t)}\left(\mu_1=\ln \lambda_1,\ldots,\mu_n=\ln \lambda_N\right)d\mu_1 \cdots d\mu_N\nn\\
\fl&=&
\frac{1}{N! \prod_{a=1}^N \Gamma^2(a)} 
\sum_{\varepsilon \in S_N}
\det_{1\leq a,b\leq N}\left[\Gamma(a+b-1) \Phi_{ab}(\lambda_{\varepsilon(b)})d\lambda_{\varepsilon(b)}
\right] , \nn\\
\fl&&
\label{PPhi}
\eea
where
\begin{equation}
\Phi_{ab}(\lambda) = \frac{1}\lambda f_{ab}(\ln \lambda) \ .
\end{equation}
For large $t$ when $f_{ab}(\mu)$ is 
approximated by 
normal distributions, 
$\Phi_{ab}(\lambda)$ can be approximated by 
 log-normal distributions.
Otherwise everything works exactly in the same way as for Lyapunov exponents. 
In particular, when $t$ is large enough to neglect the $1/t$ corrections, 
we obtain the counterpart of eq. (\ref{leading})
\begin{equation}
\label{leading_esv}
P_N^{(t)}(\lambda_1,\ldots,\lambda_N) \overset{t\gg1}{\approx}
\frac{1}{N!} \mathrm{per}_{1\le a,b\le N} \left[\Phi_b(\mu_a)\right]
\end{equation}
with 
\begin{equation}
\Phi_{b}(\lambda) \equiv \frac{1}{\sqrt{2\pi\sigma^2_{b}}\ \lambda} 
\exp\left( -\frac{(\ln \lambda - m_{b})^2}{2\sigma_{b}^2}\right)
\end{equation}
and $m_b$, $\sigma_b^2$ are given by eq. (\ref{single_index}). The functions
$\Phi_b(\lambda)$ have maxima at $\exp[\psi(b)/2]$.
The density of incremental
singular values is given by the normalized sum
\begin{eqnarray}
\fl\rho_{N}(\lambda) = \la^{-1}\rho_N(\mu=\ln(\la))=\frac{1}{N} \sum_{b=1}^N \Phi_b(\lambda)
=\frac{1}{N} \sum_{b=1}^N 
\frac{1}{\sqrt{2\pi\sigma^2_{b}}\ \lambda} 
\exp\left( -\frac{(\ln \lambda - m_{b})^2}{2\sigma_{b}^2}\right) ,\nn\\
\fl\label{rhoincr}
\end{eqnarray} 
in analogy to eq. (\ref{one-point}). Again this turns into a sum of delta functions 
in the limit $t\rightarrow \infty$,
\be
\lim_{t\to\infty}\rho_{N}(\lambda)=\frac1N \sum_{b=1}^N \delta\left(\lambda-\e^{\psi(b)/2}\right).
\ee
We have tested this prediction against Monte-Carlo simulations for finite size
systems. In Fig. \ref{esv_histograms} we show histograms of incremental singular values calculated
analytically and numerically.
We see that the log-normal functions provide a very good approximation
to the actual shapes.

\section{Lyapunov exponents from the moduli of complex eigenvalues }\label{RE_fs}

Rather than using singular values, the complex eigenvalues, $Z_n(t)=R_n(t)\e^{\imath\varphi_n(t)}$, $n=1,\ldots,N$, are an alternative way to characterize the spectral properties of the matrix $\Pi(t)$, see eq.~(\ref{Pi}). In general the singular values and the moduli of complex eigenvalues are unrelated, apart from 
their product which is equal to $|\det\Pi(t)|$ and 
bounds on their Euclidean norm which result from the trace $\Tr\Pi^\dag(t)\Pi(t)$ (see eq.~\eref{inequality}), respectively.
However in the large $t$ limit, the moduli $R_n(t)$ of the complex eigenvalues will behave exactly in the same way as the singular values
$\sqrt{s_n(t)}$. In fact repeating the same construction as in section~\ref{LE_fs},  taking the $t$-th root of $R_n(t)$ will lead to the very same normal distribution, frozen at identical positions as the limiting singular values. For that reason we will use the same term Lyapunov exponent which is otherwise reserved for the singular values, only. 

We pursue a calculation similar to section~\ref{LE_fs}. Thereby we first show that all complex eigenvalues $Z_n(t)$ can be traced back to decoupled random variables apart from a trivial determinantal coupling, see subsection~\ref{sec:decoupling}. In the second step we employ the cumulant expansion to find Dirac delta functions in the leading order and Gaussian (for the corresponding Lyapunov exponents) and log-normal (for the moduli of eigenvalues) distributions in the next-to-leading order, see subsection~\ref{large-tZ}. In subsection~\ref{sec:alternative}, we present an alternative approach by first integrating over the angles $\varphi_n(t)$ and then  taking the limit $t\to\infty$. This alternative construction is also applied to the case $\beta=4$ since the analytical result for the joint probability density of the complex eigenvalues is known~\cite{Jesper,IpsKie,ais} for this case as well.

\subsection{Reduction to ``decoupled'' random variables}\label{sec:decoupling}

The definition~(\ref{Lydef}) of Lyapunov exponents requires to take the $t$-th root and the logarithm of the positive singular values. However, for complex variables this is not a unique procedure. If one takes for example the root $Z^{1/t}$, the question arises which of the $t$ roots we have to take. When choosing the primary root the resulting spectrum will be mapped onto a circular sector of the angle $2\pi/t$ which eventually shrinks to the positive semi-axis in the limit $t\to\infty$. Another alternative choice is to take the root of the moduli of the eigenvalues only, i.e.
\be
\label{root}
Z_n(t)=R_n(t)\e^{\imath\varphi_n(t)}\ \longrightarrow\ R_n^{1/t}(t)\e^{\imath\varphi_n(t)}.
\ee
Indeed this choice seems to be a more natural construction. When multiplying the product $\Pi(t)$ by new matrices, the angular parts $\varphi_n(t)$ of the eigenvalues will run around on circles while the radial part $R_n(t)$ will either exponentially contract or expand. Thus  it is not the angular part we have to worry about in the large $t$ limit since it stays in a compact set. It is the radial part of the eigenvalues which has to be rescaled such that the support stays fixed. Therefore we decide for the rooting~\eref{root}. We emphasize that the kind of rooting is crucial to find our results which may change for other constructions.

The definition of the Lyapunov exponents at  finite and infinite $t$ starting from the moduli of complex eigenvalues are
\be
\label{ft_LC}
\nu_n(t)\equiv \frac{\ln R_n(t)}{t}
\ee
and 
\be
\label{LyCdef}
\nu_n\equiv \lim_{t\to\infty}\frac{\ln R_n(t)}{t} .
\ee
 These definitions are the analog of those for the Lyapunov exponents corresponding to the singular values, see Eqs.~(\ref{Lydef}) and (\ref{ft_L}). Hereby recall that the variables $s_n(t)$ are the squared singular values which results in an additional prefactor $1/2$.
 
The initial point of our calculation 
is an exact expression for the joint probability distribution of the complex 
eigenvalues of the product matrix $\Pi(t)$ eq.~(\ref{Pi}) at finite $N$ and $t$, see \cite{ab,ARSS},
\begin{eqnarray}
\label{Pev}
\fl{P}_N^{(t)}(Z_1,\ldots,Z_N)d^2Z_1\cdots d^2Z_N=
\frac{d^2Z_1\cdots d^2Z_N}{N!\pi^N \prod_{a=1}^N \Gamma^{t}(a)} |\Delta_N(Z)|^2 \prod\limits_{b=1}^N G^{t,\,0}_{0,\,t}\left(\left.\mbox{}_{0,\ldots,0}^{-} \right| \, |Z_b|^2\right) ,
\end{eqnarray}
where $d^2Z_n$ is the flat measure in the complex plane.
As in the previous section we again drop the explicit $t$-dependence of all quantities.
We change to polar coordinates and employ the variables~(\ref{ft_LC}) such that
the joint-probability distribution reads
\begin{eqnarray}
\label{Pnustart}
&&P_N^{(t)}\left(\nu_1,\varphi_1,\ldots,\nu_N,\varphi_N\right) 
\prod_{n=1}^Nd\nu_n d\varphi_n\\
&=&
\frac{t^N \prod_{a=1}^N \exp[2t\nu_a]}{N!\pi^N \prod_{a=1}^N \Gamma^{t}(a)} 
|\Delta_N(\exp[t \nu+\imath\varphi])|^2 \prod\limits_{b=1}^N  G^{t,\,0}_{0,\,t}\left(\left.\mbox{}_{0,\ldots,0}^{-} \right| \, \exp[2t\nu_b]\right) d\nu_b d\varphi_b\nonumber .
\end{eqnarray}
We extend the first product by the identity $1=\e^{\imath\varphi_a}\e^{-\imath\varphi_a}$. With help of the identity 
$\left(\prod_{a=1}^N x_a\right) \Delta_N(x) = \det_{1\le a,b\le N}[x_a^b]$  we get
\begin{eqnarray}
&&\left(\prod_{a=1}^N \exp[2t\nu_a+\imath\varphi_a-\imath\varphi_a]\right) |\Delta_N(\exp[t\nu+\imath\varphi])|^2\\
\fl& =& 
\det_{1\le a,b\le N}\biggl[\exp[b(t\nu_a+\imath\varphi_a)]\biggl]
\det_{1\le a,b\le N}\biggl[\exp[b(t\nu_a-\imath\varphi_a)]\biggl] .
\end{eqnarray}
We expand one of these determinants
and repeat all steps which have led us from 
eq.~(\ref{Pmu}) to eq.~(\ref{det}). Thus we end up with
\begin{equation}
\label{PH}
\fl P_N^{(t)}\left(\nu_1,\varphi_1,\ldots,\nu_N,\varphi_N\right) =
\frac{1}{N!(2\pi)^N} 
\sum_{\varepsilon \in S_N} 
\det_{1\le a,b \le N}\left[ \frac{\e^{\imath(a-b)\varphi_{\varepsilon(b)}}}{\left[\Gamma(a)\Gamma(b)\right]^{t/2}} \tilde{F}_{ab}(\nu_{\varepsilon(b)}) 
\right],
\end{equation}
where
\begin{equation}
\label{Ftab}
\tilde{F}_{ab}(\nu) =  
2t G^{t,\,0}_{0,\,t}\left(\left.\mbox{}_{(a+b)/2,\ldots,(a+b)/2}^{-} \right| \, \exp[2t\nu]\right) .
\end{equation}
This function is angle-independent and positive semi-definite. It is the counterpart of $F_{ab}(\mu)$, cf. eq.~(\ref{Fab}).
This function can be normalized with help of eq.~\eref{Ginv},
\begin{equation}
\tilde{f}_{ab}(\nu)\equiv \frac{\tilde{F}_{ab}(\nu)}{\int \tilde{F}_{ab}(\nu') d\nu'}=\frac{\tilde{F}_{ab}(\nu)}{\Gamma^t\left[(a+b)/2\right]} \ ,
\end{equation}
which has again the interpretation of a probability density function. Then the joint probability density takes the form
\begin{eqnarray}
\fl P_N^{(t)}\left(\nu_1,\varphi_1,\ldots,\nu_N,\varphi_N\right) =
\frac{1}{N!(2\pi)^N} 
\sum_{\varepsilon \in S_N} 
\det_{1\le a,b \le N} \left[ \left(\frac{\Gamma\left((a+b)/2\right)}{\sqrt{\Gamma(a)\Gamma(b)}}\right)^t \e^{\imath(a-b)\varphi_{\varepsilon(b)}}
\tilde{f}_{ab}(\nu_{\varepsilon(b)}) 
\right].\nn\\
\fl
\label{Ph}
\end{eqnarray}
This is an exact expression for the joint probability distribution 
of the Lyapunov exponents constructed from the moduli of the complex eigenvalues for any $t\in\mathbb{N}$.

Skipping the definition of the moment generating function we directly turn to the cumulant generating function,
\begin{equation}
\label{gt}
\fl\tilde{g}_{ab}(\vartheta) \equiv \ln\left( \int_{-\infty}^{+\infty} d\nu \tilde{f}_{ab}(\nu) 
\exp(\nu \vartheta)\right) = t \ln \left( \frac{\Gamma\left[(a+b)/2+\vartheta/(2t)\right]}
{\Gamma\left[(a+b)/2\right]}\right),
\end{equation} 
in analogy to eq. (\ref{g}). The Taylor series of $\tilde{g}_{ab}$ at $\vartheta=0$ is
\begin{equation}\label{cumulant-exp}
\tilde{g}_{ab}(\vartheta) \equiv
\sum_{n=1}^\infty \frac{\vartheta^n }{n!} \tilde{\kappa}^{(n)}_{ab} =
\sum_{n=1}^\infty \frac{\vartheta^n}{n!} \frac{1}{2 (2t)^{n-1}} \psi^{(n-1)}\left(\frac{a+b}{2}\right) .
\end{equation}
The cumulants can be simply read off. In particular, the first two are equal to
\begin{equation}
\label{mtab}
\tilde{m}_{ab}\equiv\int d\nu \tilde{f}_{ab}(\nu) \nu= \tilde{\kappa}^{(1)}_{ab}= \frac{1}{2} \psi\left(\frac{a+b}{2}\right) 
\end{equation}
and
\begin{equation}
\label{stab}
\tilde{\sigma}^2_{ab} \equiv\int d\nu \tilde{f}_{ab}(\nu) \left(\nu-\tilde{m}_{ab}\right)^2= \tilde{\kappa}^{(2)}_{ab}= \frac{1}{4t} \psi'\left(\frac{a+b}{2}\right) .
\end{equation}
Again we underline that these results are exact for any $t\in\mathbb{N}$.

\subsection{Large $t$ limit}\label{large-tZ}

The cumulant expansion~\eref{cumulant-exp} determines the asymptotic large $t$
behavior of  $\tilde{f}_{ab}(\nu)$. Therefore we pursue the same idea as in subsection~\ref{large-tS} and center the single-variable  distribution $\tilde{f}_{ab}(\nu)$ and normalize its variance. After finding the Gaussian behavior in the large $t$ limit we go back to the non-standardized variables in the original problem.

We standardize the random variable $\nu$ by subtracting the mean and normalizing the variance to unity, which is again denoted by  an asterisk.
Consequently the higher order standardized cumulants scale as
$\tilde{\kappa}^{(n)}_{*ab}=  \tilde{\kappa}^{(n)}_{ab}/(\tilde{\sigma}_{ab})^n\sim t^{1-n/2}$ for large $t$ and $n\geq2$.
Eventually they vanish in the limit $t\rightarrow \infty$ and as a consequence, following the same argument as leading to eq.~(\ref{gauss}), the distributions
$\tilde{f}_{ab}(\nu)$ asymptotically become normal, i.e.
\begin{equation}
\label{tf}
\tilde{f}_{ab}(\nu) \overset{t\gg1}{\approx} \frac{1}{\sqrt{2\pi\tilde{\sigma}^2_{ab}}}
\exp\left( -\frac{(\nu-\tilde{m}_{ab})^2}{2\tilde{\sigma}_{ab}^2}\right)
\end{equation}
with $\tilde{m}_{ab}$ from eq. (\ref{mtab}) and $\tilde{\sigma}_{ab}$ from eq. (\ref{stab}).
This function is identical to the distribution of Lyapunov exponents corresponding to the singular values~(\ref{gauss}), with the difference that
the mean and the variance still depend on both matrix indices $a$ and $b$ in the leading order of the $1/t$ expansion.
Note, that for the diagonal elements $a=b$ and for large $t$
the functions $\tilde{f}_{bb}(\nu)$ are
identical to $f_b(\nu)$, i.e. $\tilde{f}_{bb}(\nu)\approx f_b(\nu)$ for $t\gg1$. Especially we have $\tilde{m}_{bb}=m_b$ and $\tilde{\sigma}_{bb}=\sigma_b$, cf. eqs.~\eref{single_index}, \eref{mtab}, and \eref{stab}.

Let us discuss the prefactors in the determinant~\eref{Ph},
\begin{equation}\label{prefactor}
  D_{ab}(t) \equiv \left(\frac{\Gamma\left[(a+b)/2\right]}{\sqrt{\Gamma(a)\Gamma(b)}}\right)^t 
\end{equation}
which become Kronecker symbols. For $a=b\geq1$ these prefactors are indeed equal to unity. For $a\neq b\geq 1$ we use the fact that the geometric average is larger than the arithmetic one, $[(a+b+2j)/\sqrt{4(a+j)(b+j)}]^t>1$ for all $j=0,1,\ldots$ We have
\begin{eqnarray}\label{prefactor-1}
    D_{ab}(t) &<& \left(\frac{\Gamma\left[(a+b)/2\right]}{\sqrt{\Gamma(a)\Gamma(b)}}\right)^t\left(\frac{a+b}{\sqrt{4ab}}\right)^t =\left(\frac{\Gamma\left[(a+b+2)/2\right]}{\sqrt{\Gamma(a+1)\Gamma(b+1)}}\right)^t\\  &<&\ldots<\lim_{j\to\infty}\left(\frac{\Gamma\left[(a+b+2j)/2\right]}{\sqrt{\Gamma(a+j)\Gamma(b+j)}}\right)^t =1.\nonumber
\end{eqnarray}
The limit can be done via Stirling's formula. Therefore the determinant eq.~(\ref{Ph}) reduces to the product of diagonal elements in the large $t$ limit.
As a consequence the dependence on the angles $\varphi_n$ completely disappears. Therefore we arrive at
\begin{eqnarray}
\fl P_N^{(t)}\left(\nu_1,\phi_1,\ldots,\nu_N,\phi_N\right) \overset{t\gg1}{\approx}
\frac{1}{N! (2\pi)^N} \sum_{\varepsilon \in S_N}
\prod_{b=1}^N\tilde{f}_{bb}\left(\nu_{\varepsilon(b)}\right)=\frac{1}{N!(2\pi)^N}\mathrm{per}_{1\le a,b\le N}
\left[f_b(\nu_a)\right].\nn\\
\fl
\label{Pnu_final}
\end{eqnarray}
Note that we employed the Gaussian approximation $f_b(\nu)$, see eq.~\eref{single_index_peak}, since the means, $\tilde{m}_{bb}$, and the variances, $\tilde{\sigma}_{bb}$, agree with those for the Lyapunov exponents constructed from the singular values. This is in hindsight our justification for giving them the same names.

\begin{figure}
\centerline{\includegraphics[width=0.5\textwidth]{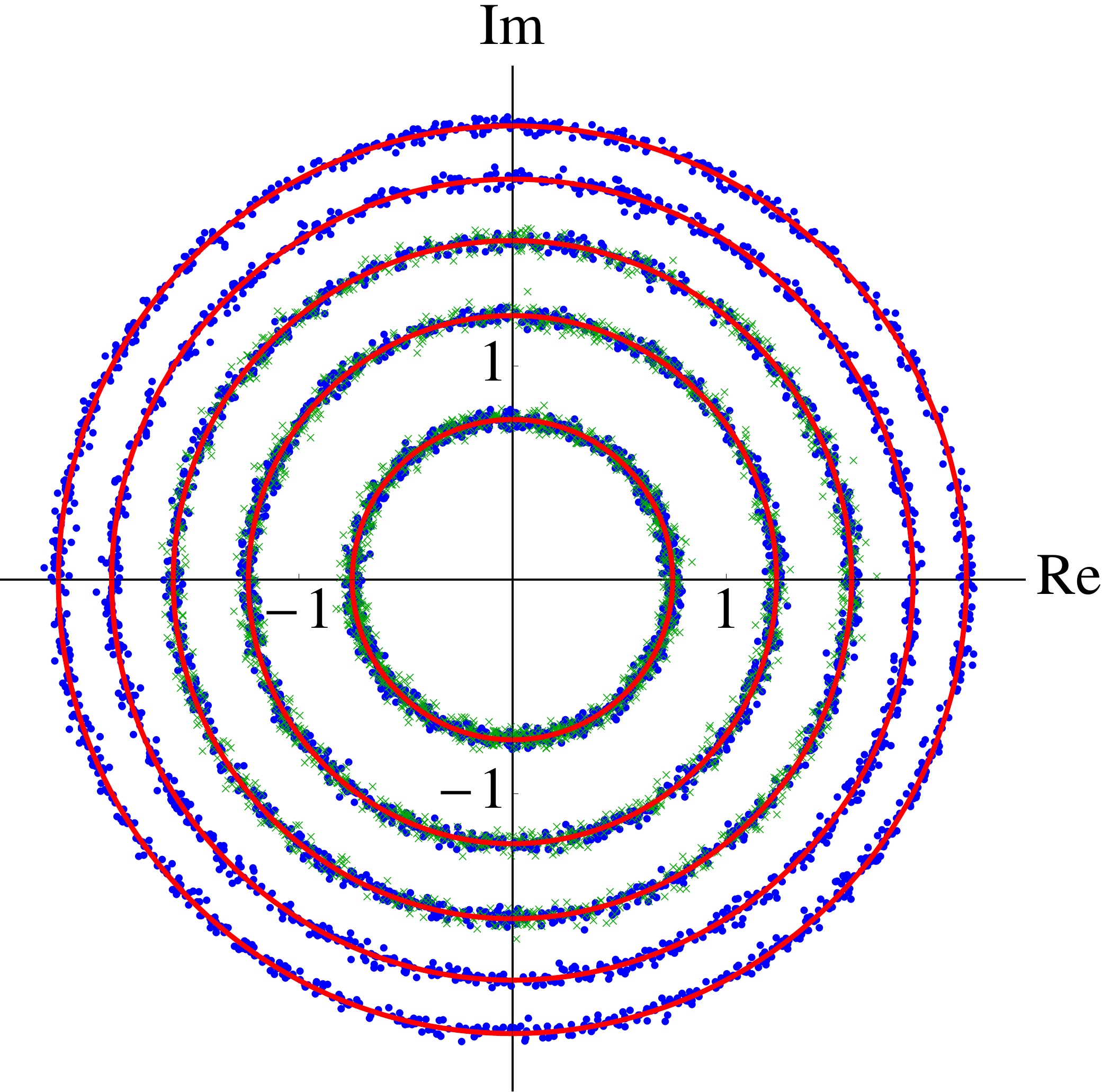}}
\caption{Scatter plot of the complex eigenvalues of the product matrices $\Pi(N=3,t=300)$ (green crosses) and  $\Pi(N=5,t=500)$ (blue dots)
derived by the rooting procedure~(\ref{root}). The plot was
generated by Monte-Carlo simulations of $1000$ product matrices for each setting. The solid red lines represent rings with
radii $\exp[\psi(b)/2]$, $b=1,\ldots,5$, given by the analytical result
in the limit $t\rightarrow \infty$. \label{rings}}
\end{figure}

Because the result~(\ref{Pnu_final}) is independent of the angles $\varphi_n$,
integrating over them yields a trivial factor $(2\pi)^N$,
\begin{eqnarray}
\fl\int_0^{2\pi} \ldots \int_0^{2\pi} d\varphi_1 \ldots d\varphi_N
P_N^{(t)}\left(\nu_1,\varphi_1,\ldots,\nu_N,\varphi_N\right)&\overset{t\gg1}{\approx}&P_N(\nu_1,\ldots,\nu_N)\nn\\
 \fl&=&\frac{1}{N!}\mathrm{per}_{1\le a,b\le N}\left[f_b(\nu_a)\right].\label{pnu_final-without-angle}
\end{eqnarray}
The resulting distribution is identical to the distribution for the Lyapunov exponents corresponding to the singular values, see eq.~(\ref{leading}). Consequently the same results apply to the density of the Lyapunov exponents obtained from the moduli of the complex eigenvalues, eq. (\ref{one-point}) and its limit as a sum of delta functions eq. (\ref{deterministic}).

It is straightforward to transform the joint probability density eq.~(\ref{Pnu_final}) back to the incremental radii $r_n \equiv \e^{\nu_n}$,
\begin{equation}
\label{res}
\fl P_N\left(\nu_1=\ln r_1,\ldots,\nu_N=\ln r_N\right)=
\frac{1}{N!
}\mathrm{per}_{1\le a,b\le N}
\left[\frac{1}{\sqrt{2\pi\sigma^2_{b}} r}
\exp\left( -\frac{(\ln r - m_{b})^2}{2\sigma_{b}^2}\right)\right].
\end{equation}
Their joint probability density
is a combination of log-normal distributions with exactly the same parameters as for the singular values
(\ref{leading_esv}). The result~(\ref{res}) implies that for large $t$ the radii $r_b$
describe narrow rings centered around the origin with their maxima at $\exp[\psi(b)/2]$, $b=1,\ldots,N$, cf. Fig.~\ref{rings}. In particular
the moduli $r_b$ have log-normal distributions and
the phases $\varphi_b$ are independent and uniformly distributed.
The determinantal repulsion between complex eigenvalues is completely lost
since they are radially separated. As a consequence the
angular degrees of freedom cease to interact and become
independent of each other in the limit $t\to\infty$.

\begin{figure}
\centerline{\includegraphics[width=0.9\textwidth]{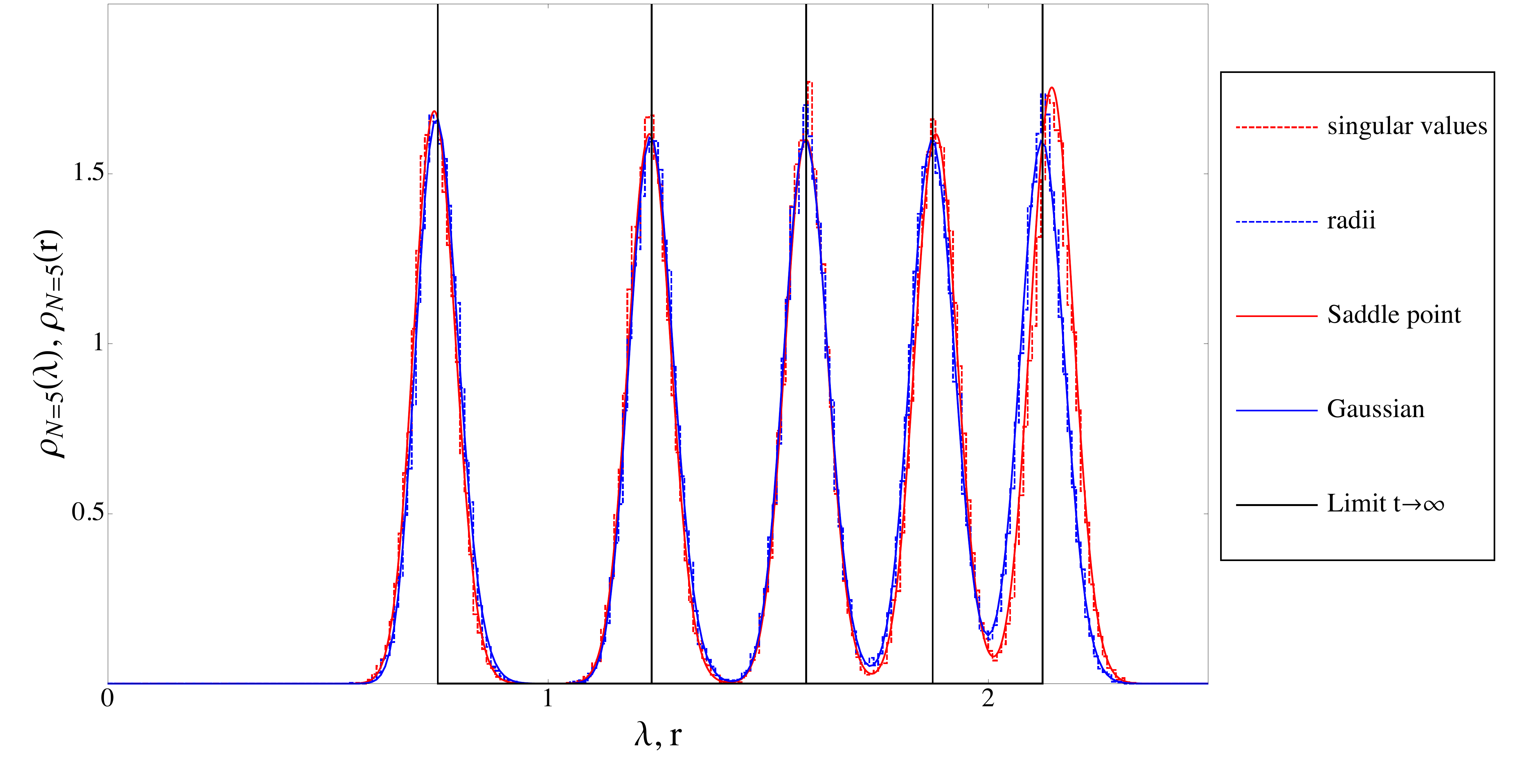}}
\caption{The histograms  show the distribution of the incremental singular values (red dashed histogram) and of the incremental radii of the complex eigenvalues (blue dashed histogram) of products of $t=100$ $5\times5$ complex Ginibre matrices generated by Monte Carlo simulations (ensemble size = $10000$ product matrices). The distribution of the radii are well approximated by the analytical result~\eref{res} (blue curve) while the corresponding saddle point approximation~\eref{Gaussapprox} for the incremental singular values (red curve) is needed for a better agreement for higher singular values. For the smallest radii and singular values all distributions perfectly agree. The positions of the limiting result $\exp[\psi(b)/2]$, $b=1,\ldots,5$, are shown by vertical lines.\label{ev_sv}}
\end{figure}

Indeed also for the results~\eref{Pnu_final} and \eref{res} we can investigate the $1/t$ correction, in particular we can apply a saddle point approximation similar to eq.~\eref{Gaussapprox}. However the Monte Carlo simulations performed show already a perfect agreement with the Gaussian approximation, see Fig.~\ref{ev_sv}. The reason is the prefactor~\eref{prefactor} in front of the single variable distributions $\tilde{f}_{ab}(\nu)$ which additionally suppresses the level repulsion. This behavior is much stronger than for the incremental singular values. Nevertheless, both distributions, the one for the radii and the singular values, will eventually agree, as it can be seen for the smallest radii and singular values in Fig.~\ref{ev_sv}.

\subsection{An alternative approach}\label{sec:alternative}

We close this section by offering a short-cut from the joint density eq. (\ref{Pnustart}) to the final result eq. (\ref{Pnu_final}).
 Once all angles are integrated out the moduli of the complex eigenvalues $Z_n$ of the product of Ginibre matrices immediately become independent random variables, see Refs.~\cite{as,ais} and for a general discussion Ref.~\cite{hkpvBook}. These integrations  can be already performed for the distribution~(\ref{Pnustart}) such that we immediately arrive at
\be
\fl\int_0^{2\pi}\prod_{n=1}^Nd\phi_n
P_N^{(t)}\left(\nu_1,\phi_1,\ldots,\nu_N,\phi_N\right)
=\frac{1}{N!}\mathrm{per}_{1\le a,b\le N}
\left[\frac{2t G^{t,\,0}_{0,\,t}\left(\left.\mbox{}_{a,\ldots,a}^{-} \right| \, \exp[2t\nu_b]\right)}{\Gamma^t(b)}
\right].
\ee
This result is still exact for finite $t$. The application of the asymptotic limit of the Meijer G-function~(\ref{gauss}) immediately leads to the following answer
\begin{eqnarray}
\fl\int_0^{2\pi}\prod_{n=1}^Nd\phi_n
P_N^{(t)}\left(\nu_1,\phi_1,\ldots,\nu_N,\phi_N\right)
\overset{t\gg1}{\approx}\frac{1}{N!}\mathrm{per}_{1\le a,b\le N}\left[
 \frac{1}{\sqrt{2\pi\sigma^2_{b}}}
\exp\left( -\frac{(\nu_a-m_{b})^2}{2\sigma_{b}^2}\right)
\right],\nn\\
\fl
\end{eqnarray}
which is identical to eq.~\eref{pnu_final-without-angle}. The parameters for the mean and variance are given in eq.~(\ref{single_index}). Let us emphasize again that the loss of angular dependence also directly results from the large $t$ limit.

Let us ask at this point about the situation for general Dyson index $\beta=1,2,4$. The integration over all angles is a non-trivial task in the case $\beta=1$ though it was shown in Ref.~\cite{f2} that in the large $t$ limit all eigenvalues become real with probability $+1$, and in Ref.~\cite{IpsKie} an expression for the joint probability density was derived for an arbitrary isotropic weight.

For $\beta=4$ the situation is much easier. Not only explicit expressions for the joint probability densities of quaternionic Ginibre matrices~\cite{Jesper}  and of general isotropic weight~\cite{IpsKie} were derived, also the integral over the angles was done \cite{ais}. Performing these integrals also leads to a permanent, which reads for Ginibre matrices
\be
\fl\int_0^{2\pi}\prod_{n=1}^Nd\phi_n
P_N^{(t,\ \beta=4)}\left(\nu_1,\phi_1,\ldots,\nu_N,\phi_N\right)
=\frac{1}{N!}\mathrm{per}_{1\le c,d\le N}
\left[\frac{2t G^{t,\,0}_{0,\,t}\left(\left.\mbox{}_{2c,\ldots,2c}^{-} \right| \, \exp[2t\nu_d]\right)}{\Gamma^t(2b)}
\right].
\ee
The asymptotic limit~(\ref{gauss}) of the Meijer G-function still applies, one has to set $a=1$, $d=b$ and $b=2c$ in eq.~\eref{gauss}. This yields for the Lyapunov exponents constructed from the moduli of the complex eigenvalues
\begin{eqnarray}
 &&\int_0^{2\pi}\prod_{n=1}^Nd\phi_n
P_N^{(t,\ \beta=4)}\left(\nu_1,\phi_1,\ldots,\nu_N,\phi_N\right)\nn\\
 &\overset{t\gg1}{\approx}&\frac{1}{N!}\mathrm{per}_{1\le c,d\le N}\left[\frac{1}{\sqrt{2\pi\sigma^2_{2c}}}\exp\left( -\frac{(\nu_d-m_{2c})^2}{2\sigma_{2c}^2}\right)\right].\label{beta4result}
\end{eqnarray}
Note the similarity to eq.~\eref{pnu_final-without-angle} although  the product now consists of quaternion matrices, only. Nevertheless, we have to be careful when interpreting this result as a hint that the final level statistics for $\beta=4$ become, apart from a factor $2$ in the indices, identical to the ones for $\beta=2$. The scatter plots in Fig.~\ref{rings2} show that  the eigenvalues are by far not uniformly distributed along the rings. Thus the angular distribution will be non-trivial for $\beta=4$.

When taking the exact limit $t\to\infty$ of eq.~\eref{beta4result} the Gaussian functions convert to into Dirac delta functions at the positions $\nu=\psi(2c)/2$, $c=1,\ldots,N$. These positions were already found by Kargin~\cite{k2} for the Lyapunov exponents from singular values for the product of quaternionic Ginibre matrices.

Indeed it would be nice to find also the finite $t$ corrections to this limit for the singular values for $\beta=1,4$. However the group integrals involved in this problem prevent an explicit expression for the joint probability density, see \cite{akw,aik} for comparison to the approach applied to the case $\beta=2$. Nonetheless we conjecture that the Lyapunov exponents from singular values and moduli of complex eigenvalues should again coincide as for $\beta=2$. This conjecture is at least confirmed by Monte Carlo simulations, see Fig.~\ref{rings2}, as well as by a direct analysis of $2\times2$ matrices, see subsection~\ref{sec:2x2}.

\section{Large {\boldmath $N$} limit} \label{lnl}

Let us take the limit $N\to\infty$, too. In particular, we ask the question whether the limits $t\to\infty$ and $N\to\infty$ commute. This question is at the heart of understanding both kinds of limits. In particular one can consider the local spectral statistics as well as the global one.

Let us stick first to the global statistics and the situation where we take $t\to\infty$ first. For this purpose two important remarks  concerning the limit $N\to\infty$ are in order.
The complex eigenvalues of an $N\times N$ Ginibre matrix $X_j$ are scattered on a disk
of radius which grows approximately as $\sqrt{N}$. 
Therefore we have to fix the support by rescaling the matrices,
\begin{equation}
X_{*j} = \frac{X_j}{\sqrt{N}} ,\ j=1,\ldots,t,
\end{equation}
to find a proper limit for the macroscopic level density in the limit $N\to\infty$. Then the spacing between the complex eigenvalues as well as between the singular values tends to zero and the spectral distributions become continuous functions for $N\to\infty$.
In particular, the limiting eigenvalue distribution of rescaled Ginibre matrices 
is given by a uniform density on the unit disk centered at the origin 
of the complex plane which is the so-called circular law. Exactly this circular law is also found for a product of complex Ginibre matrices after taking the root of the radii for $t$ fixed and $N\to\infty$, cf. Refs.~\cite{k2,bjw}.

After rescaling the moduli of the complex eigenvalues are on average smaller
or equal to unity. Thereby the corresponding evolution $\vec{x}_{t+1} = X_{*t} \vec{x}_t$
is contractive and hence the Lyapunov
exponents are expected to be non-positive. Because the evolution is linear
the incremental singular values (or radii) rescale as
$\lambda_{*n} = \lambda_n/\sqrt{N}$.  Quantities corresponding to this normalization are denoted
by an asterisk in this section.

The rescaling results in a trivial shift for the Lyapunov exponents, i.e.
\begin{equation}
\hat{\mu}_{*b} = \frac{1}{2} \left(\psi(b) - \ln N \right) ,\ b=1,\ldots,N.
\end{equation}
The smallest Lyapunov exponent 
is approximately equal to $\hat{\mu}_{*1} \approx -1/2 \ln N$ for $N\gg1$
and the largest one is 
\begin{equation} 
\hat{\mu}_{*N} = \frac{1}{2}\left(\psi(N)-\ln N\right)\overset{N\gg1}{\approx} -\frac{1}{4N} .
\end{equation}
Therefore all Lyapunov exponents are negative and for $N\rightarrow \infty$ the spectrum extends 
from $-\infty$ to $0$. The probability that a randomly chosen exponent
$\mu'_*$ is less or equal to $\hat{\mu}_{*b}$ is
\be
\mathrm{Prob}(\mu'_* \le \hat{\mu}_{*b}) = \frac{b}{N} .
\ee
Choosing the rescaled variable $x=b/N\in]0,1]$ this probability reads
\begin{equation}
\mathrm{Prob}\left(\mu'_* \le 
\frac{1}{2} \left(\psi(Nx) - \ln N \right)\right) = x.
\end{equation}
In the limit $N\rightarrow \infty$ 
this variable becomes a continuous variable $x\in]0,1]$.
Moreover, for any fixed $\mu_*$
we can approximate $\psi(Nx) \approx \ln(Nx)+O(1/N)$ such that we have
\begin{equation}
\mathrm{Prob}\left(\mu'_* \le \frac{\ln(x)}{2} \right) 
\overset{N\gg1}{\approx} \int_{-\infty}^{\ln(x)/2} \rho_*(\mu'_*) d\mu'_*  = x.
\end{equation}
Here 
\be
\rho_*(\mu_*) \equiv\lim_{N\to\infty} \rho_{*N}(\mu_*)
\ee
is the limiting density of Lyapunov exponents for the
product of independent normalized Ginibre matrices $X_{*j}$ from eq. (\ref{deterministic}). The last equation 
can be easily solved for $\rho_{*}(\mu_*)$,
\begin{equation}
\label{rho_Lyap}
\rho_{\mu*}(\mu_*) = 2 \e^{2\mu_*}  , \quad \mu_* \ \le 0  .
\end{equation}
Changing from Lyapunov exponents to incremental singular values 
$\lambda_{*b} = \e^{\mu_{*b}}$, we obtain
\begin{equation}
\label{triangular}
\rho_{*}(\mu_*)d\mu_*=
\rho_{*}(\lambda_*)d\la_* = 2 \lambda_* d\la_* , \quad \lambda_*\in [0,1]  .
\end{equation}
This is the celebrated triangular law first derived by Newman \cite{n,ni}.

Obviously one can repeat
exactly the same calculations starting from the moduli of complex eigenvalues and obtains the same results, replacing $\mu_*\to\nu_*$ and $\la_*\to r_*$. We note in passing that the triangular distribution of  
incremental radii is identical to
the limiting radial distribution of the complex eigenvalues of normalized 
Ginibre matrices $X_*/\sqrt{N}$, $N\rightarrow \infty$, 
which is given by the uniform distribution on the complex unit disk. Here the linear behavior is nothing more than the Jacobian resulting from the choice of polar coordinates.

\begin{figure}
\centerline{\includegraphics[width=0.9\textwidth]{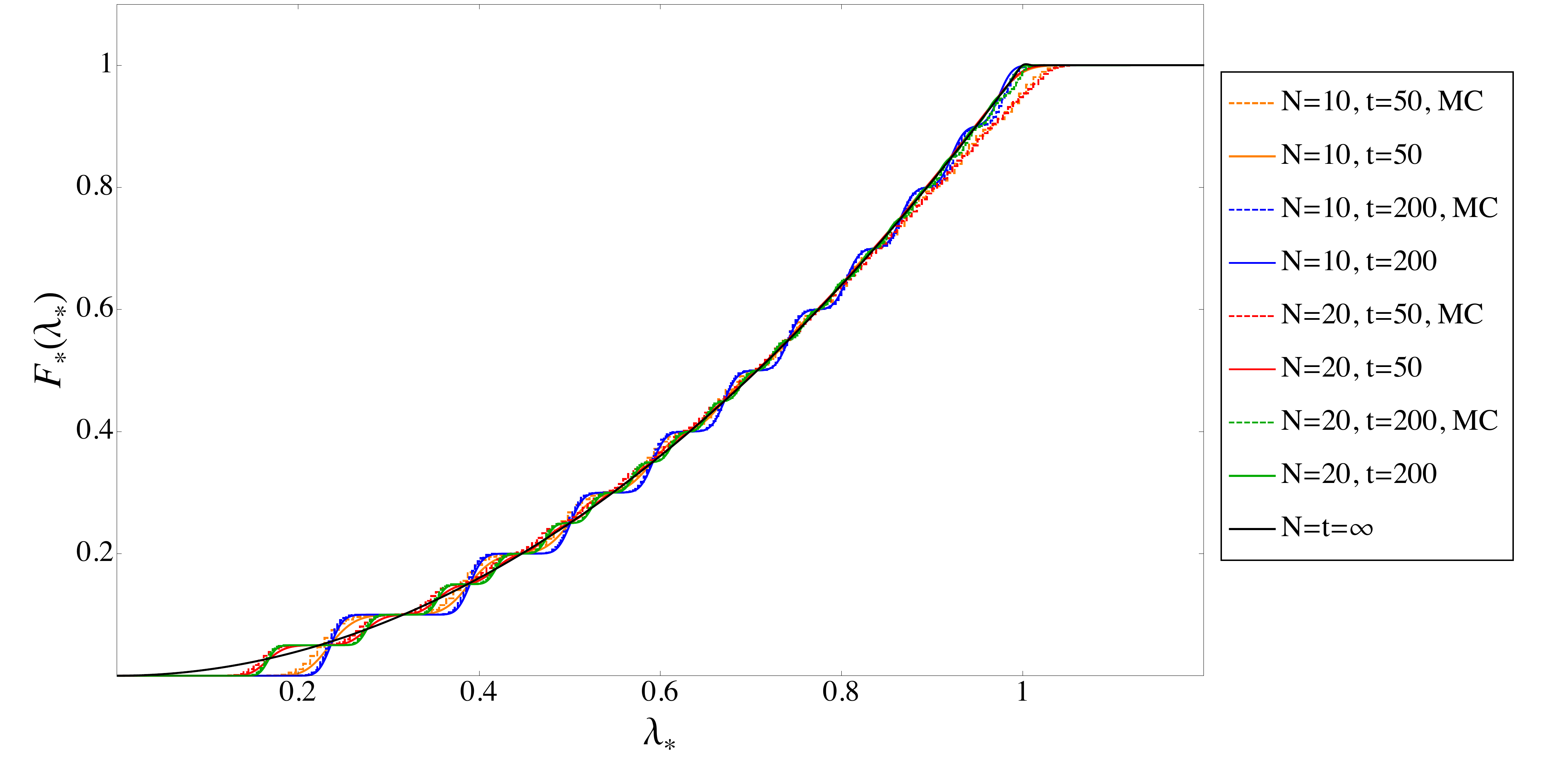}}
\caption{The analytical results (solid curves) for the cumulative distribution for the incremental
singular values $F_{*}(\lambda_*)$ are compared to Monte Carlo simulations  (dashed histograms) for varying matrix dimension $N$ and varying numbers of matrices $t$ in the product $\Pi(t)$. The black solid curve is the $N, t\to\infty$ result.\label{FF}}
\end{figure}

It is instructive to examine the convergence
of the finite $N$ distribution to the limiting triangular law. 
The cumulative distribution for the triangular law defined as 
\begin{equation}
F_{*}(\lambda_*) = \left\{
\begin{array}{ll}
\int_0^{\lambda_*} d \lambda'_* \rho_{*}(\lambda'_*) = \lambda_*^{2} 
&,\  \lambda_* \in [0,1], \\
1 &,\  \lambda_*\geq1 \\
\end{array}
\right.
\end{equation}
is trivially obtained. It is the probability to find a singular value smaller than $\lambda_*$. For finite $N$ 
(and $t \to \infty$) the cumulative distribution is just the  
counting function
\begin{equation}
F_{*N}(\lambda_*) = \frac{1}{N} \sum_{n=1}^N 
\Theta\left(\lambda_*-\frac{\exp[\psi(n)/2]}{\sqrt{N}}\right) ,
\end{equation}
with $\lim_{N\to\infty}F_{*N}(x)=F_*(x)$.
We show the evolution of the shape of this staircase function in $N$ and $t$ in Fig. \ref{FF}.

Let us study if the limits $t\rightarrow \infty$
and $N\rightarrow \infty$ commute. Therefore we consider the moments of the density of the singular values which are for  the triangular law
\begin{equation}
\label{mom_triang}
\lim_{N\to\infty}\lim_{t\to\infty}
\langle \lambda_*^n\rangle_* 
=\int_0^1 d \lambda_* \rho_{*}(\lambda_*) \lambda_*^n = \frac{2}{n+2}\ {\rm for\ all}\ n>-1.
\end{equation}
Recall that this law is obtained by taking first the limit $t\rightarrow \infty$
and then $N\rightarrow \infty$. Let us invert this order. The moments of the singular value distribution of the product 
of $t$ normalized Ginibre for $N\rightarrow \infty$ 
is equal to the Fuss-Catalan numbers \cite{Karol}
\begin{equation}
\lim_{N\to\infty}
\langle s_*^k(t)\rangle_{*} =
\frac{1}{tk+1}\left(\begin{array}{c} (t+1)k \\ k \end{array}\right)\ {\rm for\ all}\ k>-\frac{1}{t+1}.
\end{equation}
We choose $k = n/(2t)$ while keeping $n$ fixed and sending $t$ to infinity. 
Changing the integration variable
from singular values to their roots, $\lambda = s_*^{1/(2t)}$, we get
$s_*^k(t) =\lambda_*^n(t)$.
The binomial symbol on the right hand side tends to unity for 
$t\rightarrow \infty$, and the prefactor to $2/(n+2)$. Combining everything we have
\begin{equation}\label{limcom}
\lim_{t\to\infty}
\lim_{N\to\infty}
\langle \lambda_*^n(t)\rangle_{*}=\lim_{t\to\infty}
\lim_{N\to\infty}
\langle s_*^k(t)\rangle_{*}
= \frac{2}{n+2}=\lim_{N\to\infty}\lim_{t\to\infty}
\langle \lambda_*^n\rangle_* .
\end{equation}
We see that indeed the limits
$t\rightarrow\infty$ and $N\rightarrow \infty$ commute. 
To have an idea how the limiting shape of the distribution is
approached when $t$ and $N$ increase we plot in Fig. \ref{FF} 
the cumulative distribution for a collection of systems
with finite $t$ and $N$, showing both analytic and Monte-Carlo results.

To conclude this section we can ask if the commutativity of the two limits carries over to the local statistics as well. When taking first the limit $N\to\infty$ it was shown~\cite{ab} that the level statistics in the bulk and at the soft edge follow the universal results~\cite{ABKT} for complex Ginibre matrices. Especially the level spacing distribution in the bulk behaves for small spacing $\Delta r$ as $P(\Delta r) dr\approx \Delta r^3 d\Delta r\propto \Delta r^2 d\Delta r^2$, see Refs.~\cite{GHS,abps}. These results are independent of $t$ and, hence, will also not change when taking the limit $t\to\infty$ afterwards. When reversing the two limits, in particular when first taking the limit $t\to\infty$ and then the limit $N\to\infty$, we will find the statistics of the harmonic oscillator. This can be realized after unfolding the level spacing distribution of the incremental radii, i.e. $r_*\to r_*^2$, the level spacing distribution at finite $N$ but $t=\infty$ is
\begin{eqnarray}\label{levelspacing}
 P_N(\Delta r_*^2)&=&\frac{1}{N-1}\sum_{j=1}^{N-1}\delta\left(\Delta r_*^2-\exp\left[\psi(j+1)\right]+\exp\left[\psi(j)\right]\right)\\
 &=&\frac{1}{N-1}\sum_{j=1}^{N-1}\delta\left(\Delta r_*^2-\exp\left[\psi(j)\right]\left(\exp\left[\frac{1}{j}\right]-1\right)\right).\nn
\end{eqnarray}
In the limit $N\to\infty$ the variable $x=j/N$ becomes continuous and the sum can be approximated by an integral such that
\begin{eqnarray}\label{levelspacing2}
 \fl P(\Delta r_*^2)&\equiv&\lim_{N\to\infty}P_N(\Delta r_*^2)\\
 \fl&=&\lim_{N\to\infty}\int_0^1 dx\delta\left(\Delta r_*^2-\exp\left[\psi(Nx)\right]\left(\exp\left[\frac{1}{Nx}\right]-1\right)\right)=\delta(\Delta r_*^2-1)\nn
\end{eqnarray}
which is the one of an harmonic oscillator. This result is far away from the unfolded level spacing distribution of the Ginibre ensemble which has a linear slope in $\Delta r^2$, $P(\Delta r) dr\equiv P(\Delta r^2)d\Delta r^2\approx \Delta r^2d\Delta r^2$, for small spacing $\Delta r\ll1$, see Refs.~\cite{GHS,abps}.  Therefore on the local scale the two limits do not commute in contrast to the global scale, cf. eq.~\eref{limcom}. The same argument is expected to be true for the local statistics of the incremental singular values.

Since the two limits commute on the global scale while they do not commute on the local one, we claim that there should be a non-trivial double scaling limit where new results should show up. In particular we expect a mesoscopic scale of the spectrum which may also show  a new kind of universal statistics.

\section{Isotropic evolution with arbitrary weights} \label{ie}

So far we have discussed the evolution~(\ref{evolution})
driven by independent Ginibre matrices. An important property of  this random matrix ensemble is its isotropic nature, meaning that it is invariant under bi-unitary transformations, $d\mu (X_*) = d\mu (U X_* V^{-1})$,
with respect to the right and left multiplication of any unitary matrices $U,V\in\U(N)$. We want to generalize our discussion to more general isotropic random matrix ensembles, particularly to non-Gaussian weights. For this purpose we recall Newman's argument \cite{n} to find the Lyapunov exponents constructed from the singular values in subsection~\ref{sec:Newman}. In subsection~\ref{sec:2x2} we discuss why the Lyapunov exponents corresponding to the radii of the complex eigenvalues agree with those of the singular values for products of $2\times2$ complex matrices identically drawn from an arbitrary isotropic weight. Moreover we briefly discuss the extension of this argument to arbitrary dimension $N$ and arbitrary Dyson index $\beta=1,2,4$.

\subsection{Newman's argument for the singular values}\label{sec:Newman}

Let us recall a general argument given by Newman \cite{n},
which can be applied to an arbitrary isotropic evolution. It says that in the large $t$ 
limit the Lyapunov exponents become deterministic. This behavior is related to some kind of self-averaging different from the one discussed in \cite{bns,bls}. 

Newman's argument is based on a particular definition of the Lyapunov exponents constructed from the singular values. The sum of $k$ largest Lyapunov exponents is given by
\begin{equation}
\label{newman1}
 \Sigma_{k}( t)\equiv\hat{\mu}_N( t) + \ldots + \hat{\mu}_{N-k+1} ( t)=   
\max_{A \in\mathbb{C}^{N\times k}}  \frac{1}{2t}
\ln \frac{\det A^\dagger \Pi^\dagger(t) \Pi(t) A}{\det A^\dagger A} ,
\end{equation}
where the maximum is taken over all complex $N\times k$ matrices $A$ whose singular values do not vanish, i.e. $\det A^\dagger A\neq0$. We denote the average of an observable $O(\Pi(t))$ by
\begin{equation}\label{meanvalue}
 \langle O(\Pi(t))\rangle_t=\int d\mu (X_1)\cdots d\mu (X_t) O(\Pi(t)).
\end{equation}
Then Newman's argument is equivalent to the fact that for any integrable test function $f$ depending on the vector $\Sigma(t)=(\Sigma_{1}( t),\ldots,\Sigma_{N}( t))$ we have
\begin{eqnarray}\label{newman2}
 \lim_{t\to\infty}\langle f(\Sigma(t))\rangle_t=f(\langle\Sigma(1)\rangle_1),
\end{eqnarray}
where on the right hand side we average over a single matrix ($t=1$), only.

The idea to prove the claim~\eref{newman2} is to introduce a telescopic product in the definition~\eref{newman1},
\begin{eqnarray}\label{newman3}
\Sigma_{k}( t)&=&   \max_{A \in \mathbb{C}^{N\times k}} \left\{\frac{1}{2t} \ln \left(\prod\limits_{j=1}^t
\frac{\det A^\dagger \Pi^\dagger(j) \Pi(j) A}
{\det A^\dagger \Pi^\dagger(j-1) \Pi(j-1) A} \right)\right\}\\
&=&\max_{A \in \mathbb{C}^{N\times k}} \left\{\frac{1}{2t} \sum\limits_{j=1}^t\ln \left(
\frac{\det A^\dagger_j X_j^\dagger X_j A_j}
{\det A^\dagger_j A_j} \right)\right\}\nn
\end{eqnarray} 
with $A_j=\Pi(j-1) A$ and $\Pi(0)$ the $N$-dimensional  identity matrix. Note that the sum cannot be simply pushed through the operation ``$\max$'' since the matrices $A_j$, $j=1,\ldots,N$, depend on each other. Exactly at this point the isotropy of the weight becomes important. With the help of the average one can show that
\begin{eqnarray}\label{newman4}
 \fl\langle f(\Sigma(t))\rangle_t&=&\int d\mu (X_1)\cdots d\mu (X_t) f\left(\max_{A \in \mathbb{C}^{N\times k}} \left\{\frac{1}{2t} \sum\limits_{j=1}^t\ln \left(\frac{\det A^\dagger_j X_j^\dagger X_j A_j}{\det A^\dagger_j A_j} \right)\right\}\right)\\
 \fl&=&\int d\mu (X_1)\cdots d\mu (X_t) f\left(\frac{1}{2t} \sum\limits_{j=1}^t\ln \left(\det P_k^\dagger X_j^\dagger X_j P_k \right)\right)\\
 \fl&=&\int d\mu (X_1)\cdots d\mu (X_t) f\left(\frac{1}{2t} \sum\limits_{j=1}^t\max_{A \in \mathbb{C}^{N\times k}} \left\{\ln \left(\frac{\det A^\dagger X_j^\dagger X_j A}{\det A^\dagger A} \right)\right\}\right).\nn
\end{eqnarray}
The reason is that $A_j$ has the singular value decomposition $A_j=U_jP_k\Lambda_jV_j$, with $U_j\in\U(N)$, $V_j\in\U(k)$, $\Lambda_j={\rm diag}(\lambda_{1j},\ldots,\lambda_{kj})\in\mathbb{R}^k_+$, and $P_k$ the matrix mapping $k$-dimensional vectors as $v=(v_1,\ldots,v_k)\in\mathbb{C}^k$ to the trivially embedded $N$-dimensional vectors $(v_1,\ldots,v_k,0,0,\ldots,0)\in\mathbb{C}^N$. The matrix $V_j$ as well as the diagonal matrix $\Lambda_j$ trivially drop out of the ratios of determinants. The matrix $U_j$ can be readily absorbed in the measure of $X_j$ due to the substitution $X_j\to X_jU_j^\dagger$ and the isotropy of the measure $d\mu(X_j)$. Thus everything only depends on  the matrices $X_j$ and on the embedding (projection) matrix $P_k$ which is independent of $A_j$. Therefore we can completely omit taking the maximum of $A=U_1\Lambda_1V_1$, cf. the second line of eq.~\eref{newman4}, and exchange the sum with the maximum. To restore the dependence on $A_j$ we substitute $X_j\to X_jU_1$ anew. Hence we find the identity~\eref{newman4}.

In the limit $t\to\infty$ the sum is equal to the average of a single random matrix because of the law of large numbers. In particular we have
\begin{equation}\label{newman5}
\hat{\mu}_N + \ldots + \hat{\mu}_{N-k+1} =  \left\langle \max_{A \in\mathbb{C}^{N\times k}}  \frac{1}{2}
\ln \frac{\det A^\dagger \Pi^\dagger(1) \Pi(1) A}{\det A^\dagger A}\right\rangle_1 
\end{equation}
From this equation one can also simply determine the
incremental singular values $\hat{\lambda}_{n} = \exp[\hat{\mu}_n]$. In the case of complex Ginibre ensembles the
result~(\ref{newman5}) yields $\hat{\mu}_n = \psi(n)/2$. In Ref.~\cite{n} this proof was given for $\beta=1$, only.

We stress that the whole line of argument also applies in the case of general Dyson index $\beta=1,2,4$. One only has to assume that the weight is invariant under  right multiplication with the groups ${\rm O}(N)$, $\U(N)$ and ${\rm USp}(2N)$, respectively, and that the first moment of the Lyapunov exponents exists. Note that we only need the invariance under right multiplication. This is the reason why introducing fixed covariance matrices in the product of matrices did not cause any problems as it was considered in Ref.~\cite{f} for $\beta=2$ and in Ref.~\cite{k2} for $\beta=1,2,4$.

\subsection{Lyapunov exponents of general isotropic $2\times2$ random matrices}\label{sec:2x2}

 The question arises if products of random matrices drawn from any isotropic ensemble lead to a collapse of the Lyapunov exponents from the singular values and from the moduli of the complex eigenvalues to one and the same distribution as it was shown in sections~\ref{LE_fs} and \ref{RE_fs}. For a product of $2\times2$ random matrices this question can be answered positively. For this purpose we consider the product matrix
\begin{equation}\label{Pi2x2}
\fl\Pi(t) =\left[\begin{array}{cc} x_{11} & x_{12} \\ x_{21} & x_{22}  \end{array}\right]= X_t X_{t-1} \ldots X_1\quad{\rm with}\quad X_j=\left[\begin{array}{cc} x^{(j)}_{11} & x^{(j)}_{12} \\ x^{(j)}_{21} & x^{(j)}_{22}  \end{array}\right]\in\mathbb{C}^{2\times2},
\end{equation}
whose random matrices are drawn from the same isotropic weight $P(X)dX=d\mu (X) = d\mu (U X V^{-1})$ with $U,V\in \U(2)$.

Let us denote the two $t$-dependent Lyapunov exponents of the singular values by $\mu_1(t)$ and $\mu_2(t)$ as defined in eq.~\eref{ft_L}.
Then Newman's argument tells us that for any integrable test function $f$ depending on $\hat{\mu}_{2}(t)=\max\{\mu_1(t),\mu_2(t)\}$ and $({\rm ln}|\det\Pi(t)|)/t=\mu_1(t)+\mu_2(t)$ we have
\begin{eqnarray}
 \lim_{t\to\infty}\langle f(\hat{\mu}_{2}(t),\mu_1(t)+\mu_2(t))\rangle_t=f(\langle\hat{\mu}_{2}(1)\rangle_1,\langle\mu_1(1)+\mu_2(1)\rangle_1).\label{singular-large-t}
\end{eqnarray}
Note that on the right hand side the average is only over a single random matrix, $\Pi(1)=X_1$.

The aim is to show that the Lyapunov exponents of the moduli of the eigenvalues $\nu_1(t)$ and $\nu_2(t)$ agree with $\mu_1(t)$ and $\mu_2(t)$ in the large $t$-limit, i.e.
\begin{eqnarray}
 \lim_{t\to\infty}\langle f(\hat{\mu}_{2}(t),\mu_1(t)+\mu_2(t))\rangle_t=f(\langle\hat{\nu}_{2}(1)\rangle_1,\langle\nu_1(1)+\nu_2(1)\rangle_1)\label{eigen-large-t}
\end{eqnarray}
with $\hat{\nu}_{2}(t)=\max\{\nu_1(t),\nu_2(t)\}$ for all integrable test functions. For this purpose we first construct an analytical relation between $\nu_{1,2}(t)$ and $\mu_{1,2}(t)$.

The isotropy allows us to absorb the unitary matrices $U_j$ resulting from the generalized Schur decomposition~\cite{IpsKie,ab,ARSS},
\begin{eqnarray}\label{decomp}
 \fl\Pi(t)=U_t\left[\begin{array}{cc} z_1 &  \Delta \\ 0 & z_2 \end{array}\right]U_t^\dagger\ {\rm with}\  X_j=U_j\left[\begin{array}{cc} z_{1j} &  \Delta_j \\ 0 & z_{2j} \end{array}\right]U_{j-1}^\dagger\ {\rm and}\ U_0=U_t.
\end{eqnarray}
The variables $z_1, z_2, \Delta\in\mathbb{C}$ depend on $z_{1j}, z_{2j}, \Delta_j\in\mathbb{C}$ via the relations
\begin{eqnarray}\label{relations}
 z_1=\prod\limits_{j=1}^t z_{1j},\ z_2=\prod\limits_{j=1}^t z_{2j},\ \Delta=\sum_{j=1}^t\left(\prod\limits_{l=1}^{j-1} z_{1l}\right)\Delta_j\left(\prod\limits_{l=j+1}^t z_{2l}\right).
\end{eqnarray}
The quantities $\hat\mu_{2}(t)$ and $\mu_1(t)+\mu_2(t)$ read in terms of the variables $z_{1/2}$ and $\Delta$ as
\begin{eqnarray}\label{singLya2x2-a}
 \fl\hat\mu_{2}(t)&=&\frac{1}{2t}{\rm ln}\left(\frac{|z_1|^2+|z_2|^2+|\Delta|^2+\sqrt{(|z_1|^2+|z_2|^2+|\Delta|^2)^2-4|z_1z_2|^2}}{2}\right),\\
 \fl\mu_1(t)+\mu_2(t)&=&\frac{1}{t}{\rm ln} |z_{1}z_{2}|=\frac{1}{t}\sum\limits_{j=1}^t ({\rm ln}|z_{1j}|+{\rm ln}|z_{2j}|)=\nu_1(t)+\nu_2(t).\label{singLya2x2-b}
\end{eqnarray}
Note that these quantities only depend on $|z_{1,2}|$ and $|\Delta|$.
After plugging these relations into the finite $t$ average over the test function $f$ and decomposing the variables $z_{1j}=R_{1j}e^{\imath\varphi_{1j}}$ and  $z_{2j}=R_{2j}e^{\imath\varphi_{2j}}$ into radial and angular parts  we obtain
\begin{eqnarray}
 \fl&&\langle f(\hat\mu_{2}\left(t\right),\mu_1(t)+\mu_2(t))\rangle_t\nonumber\\
 \fl&=&\prod\limits_{j=1}^t\left(4\int_{0}^\infty dR_{1j}dR_{2j}\int_0^{2\pi}d\varphi_{1j}d\varphi_{2j}\int_{\mathbb{C}}d^2\Delta_j \int_{\U(2)/\U^2(1)}d\chi(U_j) R_{1j}R_{2j}P\left(|z_{1j}|,|z_{2j}|, \Delta_j \right)\right)\nonumber\\
 \fl&&\times \frac{\left|\prod_{j=1}^tz_{1j}-\prod_{j=1}^tz_{2j}\right|^2}{2} f\left(\hat\mu_{2}(t),\frac{1}{t}\sum\limits_{j=1}^t ({\rm ln}|z_{1j}|+{\rm ln}|z_{2j}|)\right),\label{average-1}
\end{eqnarray}
see Ref.~\cite{IpsKie}. The factor $1/2$ results from the ordering of $z_1$ and $z_2$ which is originally included in the generalized Schur decomposition and can be lifted by taking this factor into account. The Haar measure of the co-set $\U(2)/\U^2(1)$ is denoted as $d\chi(U_j)$, $j=1,\ldots,N$. Let us stress that the isotropy of the probability density $P$ indeed allows us to absorb the dependence of $P$ on the angles of the two eigenvalues $z_1$ and $z_2$ in the integral over $\Delta$.

The integration over the phases $\e^{\imath\varphi_{1j}}$ and $\e^{\imath\varphi_{2j}}$ simplifies the integral~\eref{average-1} to
\begin{eqnarray}
 \fl&&\langle f(\hat\mu_{2}\left(t\right),\mu_1(t)+\mu_2(t))\rangle_t\nonumber\\
 \fl&=&\prod\limits_{j=1}^t\left(4\int_{0}^\infty dR_{1j}dR_{2j}\int_0^{2\pi}d\varphi_{1j}d\varphi_{2j}\int_{\mathbb{C}}d^2\Delta_j \int_{\U(2)/\U^2(1)}d\chi(U_j) R_{1j}^3R_{2j}P\left(|z_{1j}|,|z_{2j}|, \Delta_j \right)\right)\nonumber\\
 \fl&&\times  f\left(\hat\mu_{2}(t),\frac{1}{t}\sum\limits_{j=1}^t ({\rm ln}|z_{1j}|+{\rm ln}|z_{2j}|)\right),\label{average-2}
\end{eqnarray}
The collective permutation $z_{1j}\leftrightarrow z_{2j}$ employed here is legitimate. Therefore the single probability densities of the set of variables $\{z_{1j}, z_{2j},\Delta_j\}$ factorize and become statistically independent. Interestingly the average over a single set of variables $\{z_{1j}, z_{2j},\Delta_j\}$ with a fixed index $j$ is equal to the original integral over a single matrix $X_j$, i.e. eq.~\eref{average-2} also holds for $t=1$, which is quite important to find the right hand side of eq.~\eref{eigen-large-t}.

In the next step we calculate upper and lower bounds for the maximal Lyapunov exponent $\hat\mu_{2}\left(t\right)$. Looking at the relation~\eref{singLya2x2-a} it is immediate that $\hat\mu_{2}\left(t\right)$ is monotonously increasing in $|\Delta|$. Hence it is certainly true that
\begin{eqnarray}\label{esti-lower}
 \fl\hat\mu_{2}\left(t\right)&\geq&\frac{1}{2t}{\rm ln}\left(\frac{|z_1|^2+|z_2|^2+\sqrt{(|z_1|^2+|z_2|^2)^2-4|z_1z_2|^2}}{2}\right)\\
 \fl&=&\frac{1}{t}{\rm ln}\max\{|z_1|,|z_2|\}=\max\left\{\frac{1}{t}\sum_{j=1}^t{\rm ln}|z_{1j}|,\frac{1}{t}\sum_{j=1}^t{\rm ln}|z_{2j}|\right\}.\nonumber
\end{eqnarray}
Note that the sum cannot be pushed through the operation ``$\max$''.
The upper bound can be found by estimating $|\Delta|$, i.e.
\begin{eqnarray}
  \fl|\Delta|&\leq&\sum_{j=1}^t\left(\prod\limits_{l=1}^{j-1} |z_{1l}|\right)|\Delta_j|\left(\prod\limits_{l=j+1}^t |z_{2l}|\right)\leq \max_{k=1,\ldots,t}\left\{\left(\prod\limits_{l=1}^{k-1} |z_{1l}|\right)\left(\prod\limits_{l=k+1}^t |z_{2l}|\right)\right\}\sum_{j=1}^t|\Delta_j|.\nonumber\\
  \fl&&\label{esti-delta-1}
\end{eqnarray}
Because of the statistical independence of the matrices with a fixed $j$ this inequality becomes
\begin{eqnarray}\label{esti-delta-2}
  \fl\frac{1}{t}{\rm ln}|\Delta|&\leq&\max_{k=1,\ldots,t}\left\{\frac{1}{t}\left(\sum\limits_{l=1}^{k-1} {\rm ln}|z_{1l}|+\sum\limits_{l=k+1}^t {\rm ln} |z_{2l}|\right)\right\}+\frac{1}{t}{\rm ln}\left(\sum_{j=1}^t|\Delta_j|\right)\\
  \fl&\overset{t\gg1}{\approx}& \sup_{p\in]0,1[}\left\{p\langle{\rm ln}|z_{11}|\rangle_1+(1-p) \langle{\rm ln} |z_{21}|\rangle_1\right\}+\frac{1}{t}{\rm ln}\left(t\langle|\Delta_1|\rangle_1\right)\nonumber\\
  \fl&=&\langle\hat\nu_{1}(1)\rangle_1+\frac{1}{t}{\rm ln}\left(t\langle|\Delta_1|\rangle_1\right)\nn
\end{eqnarray}
  in the large $t$-limit. The latter equation results from the fact that the supremum is reached at the boundary of the interval $p\in]0,1[$ and that the moment of $|\Delta_1|$ is bounded. Therefore there is a constant $0<c<\infty$ such that
\begin{eqnarray}\label{esti-delta-3}
 |\Delta|\leq c t \exp[t \langle\hat\nu_{1}(1)\rangle_1]
\end{eqnarray}
for all  $t\in\mathbb{N}$. This inequality together with $0\leq|z_{1,2}|\leq \widetilde{c} \exp[t \langle\hat\nu_{1}(1)\rangle_1]$, where $0<\widetilde{c}<\infty$ is a second constant, yields the upper bound
\begin{eqnarray}\label{esti-upper}
 \hat\mu_{1}&\leq&\frac{1}{2t}{\rm ln}\left(2\widetilde{c}^2+c^2 t^2\right)+\langle\hat\nu_{1}(1)\rangle_1
\end{eqnarray}
for all $t\in\mathbb{N}$.

Collecting everything the bounds tell us that the large $t$ limit  is
\begin{eqnarray}\label{esti-bounds}
 \lim_{t\to \infty}\frac{1}{t}\hat\mu_{1}=\langle\hat\nu_{1}(1)\rangle_1.
\end{eqnarray}
Equation~\eref{esti-bounds} together with eq.~\eref{singLya2x2-b} prove that eq.~\eref{eigen-large-t} is indeed true. In particular it shows two things. First, the two Lyapunov exponents constructed from the moduli of the complex eigenvalues of a product of $2\times2$ matrices independently and isotropically distributed take deterministic values in the large $t$ limit. Second, the deterministic values of those Lyapunov exponents agree with those constructed from the singular values.  Both properties are true for quite general random matrix ensembles. The only additional condition apart from the isotropy is the existence of the first moments of the random variables $|\Delta_j|$ and ${\rm ln}|z_{1j,2j}|$. The existence of these moments guarantees the existence of the limits and the correctness of the calculation presented above.

Note that despite the general inequality
\begin{equation}\label{inequality}
 \Tr \Pi(t)\Pi^\dagger(t)=\sum_{j=1}^Ns_j(t)=\sum_{j=1}^N R_j^2(t)+\sum_{1\leq l<k\leq N} |\Delta_{lk}|^2\geq \sum_{j=1}^N R_j^2(t)
\end{equation}
(which is equal if and only if the matrix is normal) the agreement of both kinds of Lyapunov exponents does not immediately result in the statement that the matrix $\Pi(t)$ becomes normal in the large $t$ limit. Considering the bound~\eref{esti-delta-3} we notice that the off-diagonal element $|\Delta|$ may become exponentially large. Indeed one can easily construct such a situation by setting $|z_{1j}|,|z_{2i}|>1$ for all $i,j$. Therefore the way how we root the matrices is crucial in the large $t$ limit.

\begin{figure}
\centerline{\includegraphics[width=1\textwidth]{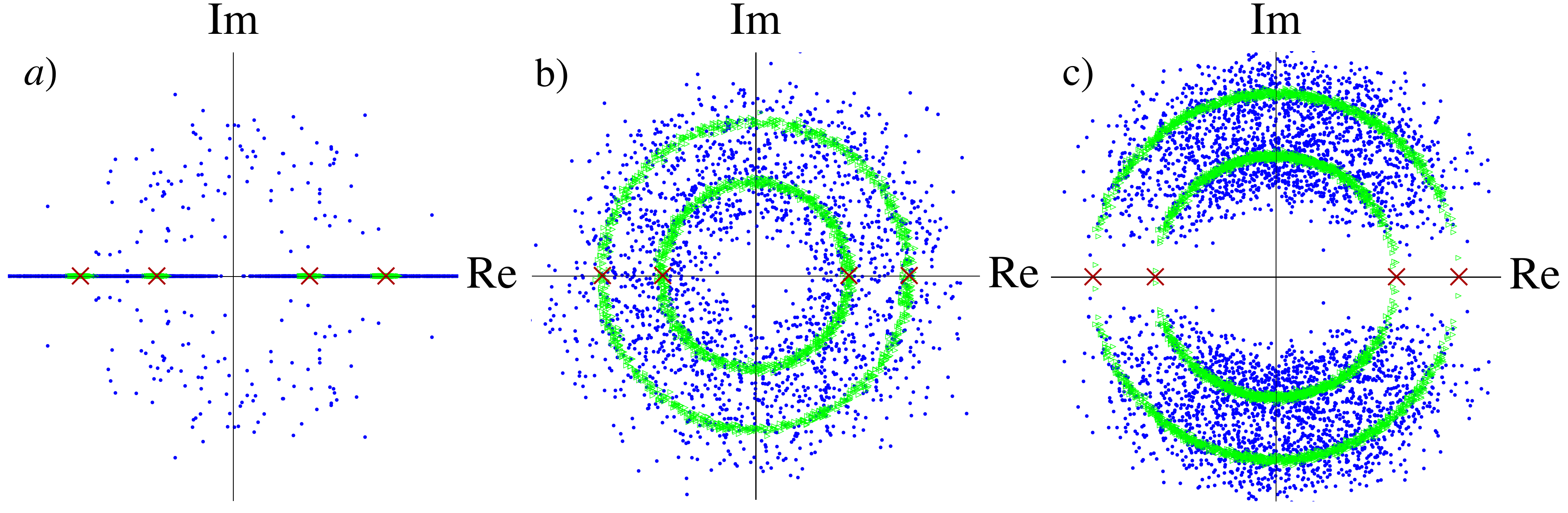}}
\caption{Scatter plots for product matrices of all three Dyson indices $\beta=1$ (a), $\beta=2$ (b), and $\beta=4$ (c). The large red crosses are the positions ($\pm\exp[\psi(\beta n/2)/2]$ with $n=1,2$) of the incremental singular values at $t\to\infty$. All three plots were generated by Monte Carlo simulations of products of Ginibre matrices for $N=2$ at $t=5$ (dark blue dots) and $t=500$ (light green triangles) drawn from an ensemble size $1000$. Note that only the case $\beta=2$ develops an angular independent spectral density while for $\beta=1$ all eigenvalues will be eventually real as proved by Forrester~\cite{f2}. For $\beta=4$ the dependence on the angle becomes non-trivial which we conjecture to be $\sin^2\varphi$.\label{rings2}}
\end{figure}

Two questions arise from our result. First, can we generalize our argument to arbitrary matrix dimension $N$? To answer this we emphasize that our calculation relies on the explicit, known relation between singular values and the components of the generalized Schur decomposition, see eqs.~\eref{singLya2x2-a} and \eref{singLya2x2-b}, which can be indeed extended to the cases $N=3,4$. Nevertheless we expect that there is a general argument. Therefore we conjecture that the Lyapunov exponents of the moduli of the complex eigenvalues and of the singular values are deterministic and agree with each other for general isotropic ensembles.

Second, can we generalize our argument to the Dyson indices $\beta=1,4$, i.e. to the product of real and quaternion Ginibre matrices? In the case $\beta=4$ and $N=2$ one can show that we find a factorization of the probability densities similar to eq.~\eref{average-2} and the same calculation can be done analogously. Therefore one can answer the question positively in this case. The situation for general $N$ is much more involved but we expect that also there the Lyapunov exponents qualitatively behave the same as in the case $\beta=2$, only their positions may change and the angles of the complex eigenvalues will not be uniformly distributed, see Fig.~\ref{rings2}.c. Regarding the distribution of the angles we expect that the density behaves as $\sin^2\varphi$. The reason is the macroscopic distance of the complex eigenvalues in the large $t$ limit such that the repulsion between the eigenvalues is suppressed. Only the repulsion of a complex conjugate pair will survive since the two eigenvalues lie on the same circle.

The case $\beta=1$ is as usual non-trivial. The matrices may have real eigenvalues as well as complex conjugate pairs, see \cite{IpsKie}. In the case $N=2$ the situation with a complex conjugate pair immediately yields that the eigenvalues condense on a fixed ring equal to the square root of the determinant of the product matrix. Newman's argument for the singular values applies to all three Dyson indices $\beta=1,2,4$ such that the modulus of the determinant becomes deterministic (it is the product of the singular values) and thus also the the moduli of the complex eigenvalue pairs. However Forrester already showed~\cite{f2} that in the large $t$ limit almost all eigenvalues will be real. The statistics of these real eigenvalues is still unclear because of the modulus of the Vandermonde determinant. Hence, the probability densities of the single matrices always remain coupled. Therefore we can conclude that the case $\beta=1$ will not yield the same result as $\beta=2$ in the angular part of the distribution. But Fig.~\ref{rings2}.a shows that the radii still seem to condense at the positions of the singular values.

\section{Conclusions} \label{conclusion}

We presented a solvable case of an isotropic time evolution with
evolution operators being independent complex $N\times N$ Ginibre matrices. 
The entire spectrum
of Lyapunov exponents, traditionally defined in terms of the singular values, was computed including their positions (which are in agreement with \cite{n,f,k1,k2}), individual and the joint probability distributions with their $1/t$ corrections in the large $t$ limit.
Surprisingly the Lyapunov exponents which can analogously be constructed for the moduli of the complex eigenvalues show exactly the same large $t$ behavior. Thereby they do not only condense on the same values as the Lyapunov exponents for the singular values but also share the same variance and normal distribution around this value. Therefore we understand this behavior as a universal property which is also expected for general isotropic weights and general Dyson index $\beta=1,2,4$.

The normal distributions with means $\psi(n)/2$ and variances $\psi'(n)/(4t)$ are the non-perturbative leading order correction to the deterministic values of the Lyapunov exponents for $t\to\infty$. They agree very well with finite $t\approx 10N$ Monte Carlo simulation for the moduli of the complex eigenvalues while for the singular values we showed that the saddle point approximation of the inverse Fourier transform of the moment generating function yields a better agreement for finite $t$. The reason is the underlying structure involved in this problem. The joint probability distributions of the singular values and of the complex eigenvalues are given by determinantal point processes reflecting the level repulsion. In the large $t$ limit this repulsion is suppressed and a permanent remains in both cases. The convergence to this result is enhanced for the eigenvalues by prefactors which are absent for the singular values. This shows that the mechanism how the singular values and the eigenvalues approach their deterministic values $\psi(n)/2$ is different. Nonetheless they share a particular  asymptotic expansion of the Meijer G-function with large index and argument which is still at the heart of taking the limit $t\to\infty$.

The limiting angular dependence is uniform for $\beta=2$. This behavior is in contrast to the case for the product of real and quaternion Ginibre matrices. In the real case all eigenvalues become real~ \cite{f2} while in the quaternion case the level density exhibits a non-trivial angular dependence. Nevertheless we claim that the radii of the eigenvalues will approach the same values as the singular values for all three Dyson indices and general isotropic random matrix ensembles in the limit $t\to\infty$. This is supported by our numerical simulations as well as by a discussion of the case $N=2$. We also considered the case $\beta=4$ for Ginibre matrices and found that the Lyapunov exponents constructed from the moduli of the complex eigenvalues indeed take the limit $\psi(2n)/2$ derived for the Lyapunov exponents corresponding to the singular values \cite{k2}. 

Moreover, we showed that the
triangular law for $N\to\infty$ can be simply interpreted as the radial distribution of the
Ginibre ensemble of the limiting circular law. Thereby we proved that the two limits $t\to\infty$ and $N\to\infty$ commute on the global scale of the spectrum of the product matrix. This commutativity is not valid anymore on the local scale. On the scale of the mean level spacing of the complex eigenvalues the limits by taking $N\to\infty$ first  and then $t\to\infty$ yield a level repulsion as found for a complex Ginibre ensemble, i.e. $P(\Delta r)d\Delta r\approx \Delta r^3d\Delta r$ for $\Delta r\ll1$, see Refs.~\cite{GHS,abps}.  Reversing this order we find the level statistics of the harmonic oscillator for the radii squared. Therefore one has to be careful on which scale of the spectrum one takes both limits. We conjecture the existence of a non-trivial scale of a double scaling limit due to this insight.

\section*{Acknowledgments}

We like to thank the SFB$|$TR12  ``Symmetries and Universality
in Mesoscopic Systems'' of the German research council DFG for partial support (G.A.).
Z.B. was supported by the Alexander von Humboldt Foundation and
the Grant DEC-2011/02/A/ST1/00119 of the National Centre of Science in Poland, and M.K. was supported by a Feodor Lynen return fellowship of the Alexander von Humboldt Foundation. We also thank Jesper R. Ipsen as well as Jens Markloff for fruitful discussions.

\begin{appendix}
\section{Some identities for Meijer G-functions}\label{appA}

Meijer G-functions are a broad class of special functions comprising most of the known special functions. 
They are defined as the inverse Mellin transform 
of certain quotients of products of gamma functions. 
We do not give their general definition, but we 
restrict ourselves to a small subclass of Meijer G-functions 
which are used in our calculations. 

We consider Meijer G-functions of the following form given by an integral~\cite{gradbook}
\begin{equation}
\label{Gdef}
G^{t,\,0}_{0,\,t}\left(\left.\mbox{}_{a_1,\ldots,a_t}^{-} \right| \, s\right)=
\int_{\mathcal{C}} \Gamma(a_1-u) \ldots \Gamma(a_t-u) s^{u}\frac{du}{2\pi \imath} ,
\end{equation}
over a contour $\mathcal{C}$ that goes from $-\imath \infty$ to $+\imath \infty$ 
leaving all poles of the Gamma functions on the right hand side.
The Mellin transform of this function is
\begin{equation}
\label{Ginv}
\int_0^\infty ds s^{u-1} G^{t,\,0}_{0,\,t}\left(\left.\mbox{}_{a_1,\ldots,a_t}^{-} \right| \, s\right) = \Gamma(a_1-u) \ldots \Gamma(a_k-u) \ .
\end{equation}
Moreover Meijer G-functions fulfill the simple but useful identity 
\begin{equation}
\label{mGp}
s^b G^{t,\,0}_{0,\,t}\left(\left.\mbox{}_{a_1,\ldots,a_t}^{-} \right| \, s\right)=
G^{t,\,0}_{0,\,t}\left(\left.\mbox{}_{b+a_1,\ldots,b+a_t}^{-} \right| \, s\right) . 
\end{equation}
which is needed several times in our calculations. This identity is 
a consequence of the shift $s^u \rightarrow s^{u+b}$ in the 
power in the integrand~(\ref{mGp}) which can be compensated by the substitution $u\rightarrow u-b$.

\section{Computation of the normalizing Hankel determinant}\label{appB}

In order to be self contained we calculate the Hankel determinant appearing in eq. (\ref{Hankel}), 
\bea
\det_{1\leq a,b\leq N}\left[\Gamma(a+b-1)\right] = \prod_{a=1}^N\Gamma^2(a) ,
\label{Hankel2}
\eea
which is a special case of a results by Normand~\cite{Normand}. We do this by 
applying Andreief's formula~\cite{Andy}
\begin{equation}
\label{andreief}
\fl\det_{1\le a,b \le N}\left[ \int dx \, \Phi_a(x) \Psi_b(x) \right]
= \frac{1}{N!} \int dx_1 \ldots dx_N 
\det_{1\le a,b \le N}\left[\Phi_a(x_b)\right]
\det_{1\le a,b \le N}\left[\Psi_a(x_b)\right].
\end{equation}
Here $\{\Phi_a(x)\}$ and $\{\Psi_a(x)\}$, $a=1,\ldots,N$ are two
sets of integrable functions of a real variable. 
 
The Gamma functions on the left hand side of (\ref{Hankel2})
can be written as
\begin{equation}
\Gamma(a+b-1) = \int_0^\infty dx \, x^{a+b-2} \exp(-x) = 
\int_0^\infty dx \, \Phi_a(x) \Psi_b(x),
\end{equation}
such that we identify $\Phi_a(x) =\Psi_a(x) = x^{a-1} \exp(-x/2)$ for $x\ge 0$, $a=1,\ldots,N$.
Andr\'eief's formula then yields
\begin{equation}
\label{Vd}
\fl\det_{1\leq a,b\leq N}\left[\Gamma(a+b-1)\right]=
\frac{1}{N!} \int dx_1\ldots dx_N 
\left(\det_{1\le a,b\le N}\left[x_b^{a-1} \exp\left(-\frac{x_b}{2}\right)\right]\right)^2.
\end{equation}
Due to the skew-symmetry of the determinant under permutations as well as its multi-linearity the rows can be linearly combined without changing its value. The idea is to combine them in such a way that after applying the Andr\'eief integral again we have to take a determinant of diagonal elements, only. The Laguerre polynomials in monic normalization, denoted by
\begin{eqnarray}\label{Laguerre}
 L_n(x)=\sum_{j=0}^n\left(\begin{array}{c} n \\ j \end{array}\right)\frac{(-1)^{n-j}n!}{j!}x^j,
\end{eqnarray}
will do the job. They are orthogonal with respect to the weight $\exp[-x]dx$, i.e.
\begin{eqnarray}\label{orthogonal}
 \int_0^\infty dx \exp[-x] L_a(x)L_b(x)=(a!)^2\delta_{ab}.
\end{eqnarray}
Therefore we have
\begin{eqnarray}\label{calc-1}
\fl\det_{1\leq a,b\leq N}\left[\Gamma(a+b-1)\right]&=&
\frac{1}{N!} \int dx_1\ldots dx_N 
\left(\det_{1\le a,b\le N}\left[L_{a-1}(x_b) \exp\left(-\frac{x_b}{2}\right)\right]\right)^2\\
\fl&=& \det_{1\le a,b\le N}\left[\int_0^\infty L_{a-1}(x)L_{b-1}(x) \exp\left(-x\right)\right]\nonumber\\
\fl&=& \prod_{a=0}^{N-1} (a!)^2.\nonumber
\end{eqnarray}
In the second line we employed eq.~(\ref{andreief}) and in the third line eq.~(\ref{orthogonal}). The last line is nothing else than the claim~\eref{Hankel2}.

In a similar way we want to calculate the cofactor of the Hankel determinant~\eref{Hankel},
\bea
C_{jl}=(-1)^{j+l}\underset{a\neq j,b\neq l}{\underset{1\leq a,b\leq N}{\det}}\left[\Gamma(a+b-1)\right],
\label{Hankel3}
\eea
which appears in eq.~\eref{Gaussapprox}. Also this determinant can be calculated via the Andr\'eief integral. For this purpose we introduce two integrals over the angles $\varphi_1$ and $\varphi_2$,
\bea
\fl C_{jl}=-\int_0^{2\pi}\frac{d\varphi_1}{2\pi}\int_0^{2\pi}\frac{d\varphi_2}{2\pi}\det\left[\begin{array}{cc} \displaystyle\left\{\int_0^\infty dx \, x^{a+b-2} \exp(-x)\right\}\underset{1\leq a,b\leq N}{\ } & \displaystyle\left\{e^{\imath(a-j)\varphi_1}\right\}\underset{1\leq a\leq N}{\ } \\ \displaystyle\left\{e^{\imath(b-l)\varphi_2}\right\}\underset{1\leq b\leq N}{\ } & 0 \end{array}\right].\nn\\
\fl\label{calc-2}
\eea
We use the same trick again by rearranging the columns and rows such that we have in the upper left block integrals over two Laguerre polynomials and thus a diagonal matrix. An expansion in this diagonal matrix yields
\bea
\fl C_{jl}=\prod_{a=0}^{N-1} (a!)^2\int_0^{2\pi}\frac{d\varphi_1}{2\pi}\int_0^{2\pi}\frac{d\varphi_2}{2\pi}\exp[\imath([1-j]\varphi_1+[1-l]\varphi_2)]\sum_{k=0}^{N-1}\frac{L_k(e^{\imath\varphi_1})L_k(e^{\imath\varphi_2})}{(k!)^2}.\nn\\
\fl\label{calc-3}
\eea
In the last step the two integrals, which factorize, can be performed and we find
\bea
\fl C_{jl}=(-1)^{j+l}\prod_{a=0}^{N-1} (a!)^2\sum_{k=0}^{N-1}\left(\frac{k!}{(j-1)!(l-1)!}\right)^2\frac{1}{\Gamma(k-j+2)\Gamma(k-l+2)}.
\label{calc-4}
\eea
Note that the function $1/\Gamma(z)$ is an entire function which is zero for negative semi-definite integers. Therefore the sum is usually smaller than the boundary shown here, i.e. its range is $k=\max\{j,l\}-1,\ldots, N-1$.

\section{Saddle point approximation of $f_{ab}(\mu)$}\label{appC}

We consider the saddle point approximation of the inverse Fourier transform of the moment generating function~\eref{mdef},
\begin{eqnarray}\label{mdefinv-1}
 f_{ab}(\mu)&=&\int_{-\imath\infty}^{+\imath\infty}\frac{d\vartheta}{2\pi\imath}\exp[-\mu\vartheta]M_{ab}(\vartheta)\\
 &=&\int_{-\imath\infty}^{+\imath\infty}\frac{d\vartheta}{2\pi\imath}\exp[-\mu\vartheta]\frac{\Gamma^{t-1}(b+\vartheta/(2t))\Gamma(a+b-1+\vartheta/(2t))}{\Gamma^{t-1}(b)\Gamma(a+b-1)}.\nonumber
\end{eqnarray}
After rescaling $\vartheta\to2t\vartheta$ the saddle point equation and its solution are
\begin{equation}\label{saddle pointeq}
 \fl\psi(b+\vartheta_b(\mu))=2\mu \Rightarrow \vartheta_b(\mu)=\int_{0}^\infty dy\Theta(2\mu- \psi(y))-b=\vartheta_0(\mu)-b,
\end{equation}
where $\Theta$ is the Heaviside function. In fact there are also other saddle points. However only the solution $\vartheta_b(\mu)=\vartheta_0(\mu)-b$ can be reached in the limit $t\to\infty$. We perform the saddle point expansion $\vartheta=\vartheta_0(\mu)-b+\imath\delta\vartheta/\sqrt{t}$ and find
\begin{eqnarray}
 \fl f_{ab}(\mu)&\overset{t\gg1}{\approx}&\frac{2\sqrt{t}\Gamma^{t-1}(\vartheta_0(\mu))\Gamma(a-1+\vartheta_0(\mu))\exp[-2t\mu(\vartheta_0(\mu)-b)]}{\Gamma^{t-1}(b)\Gamma(a+b-1)}\nonumber\\
 \fl&&\times\int_{-\infty}^{+\infty}\frac{d\delta\vartheta}{2\pi}\exp\left[-\frac{\psi'(\vartheta_0(\mu))\delta\vartheta^2}{2}\right]\nonumber\\
 \fl&=&\sqrt{\frac{2t}{\pi\psi'(\vartheta_0(\mu))}}\frac{\Gamma^{t-1}(\vartheta_0(\mu))\Gamma(a-1+\vartheta_0(\mu))}{\Gamma^{t-1}(b)\Gamma(a+b-1)}\exp[-2t\mu(\vartheta_0(\mu)-b)].\label{saddlepointexp}
\end{eqnarray}
This expression seems to factorize in a $b$ and an $a$ dependent part apart from the constant prefactor $1/\Gamma(a+b-1)$ but this is a misleading conclusion. The argument $\mu$ also depends on the index $b$ in the determinant~\eref{Pf}. Therefore the level repulsion corresponding to the determinant is still present in this particular approximation.

\end{appendix}
 

\end{document}